\documentclass[%
 amsmath,amssymb,
 aps,
]{revtex4-2}

\usepackage{subfigure}
\usepackage{graphicx}
\usepackage{amssymb}
\usepackage{amsmath}
\usepackage[dvipsnames]{xcolor}
\usepackage{float}
\usepackage{bm}
\def \bF {\pmb{F}}
\def \bH {\pmb{H}}
\def \bG {\pmb{G}}
\def \bL {\pmb{L}}
\def \bU {\pmb{U}}
\def \bx {\pmb{x}}
\def \by {\pmb{y}}
\def \bX {\pmb{X}}
\def \bY {\pmb{Y}}
\def \bz {\pmb{z}}
\def \bw {\pmb{w}}

\def \bp {\pmb{p}}
\def \bq {\pmb{q}}
\def \bu {\pmb{u}}
\def \bv {\pmb{v}}

\usepackage{graphicx}
\usepackage{dcolumn}
\usepackage{bm}
\usepackage[dvipsnames]{xcolor}
\usepackage[normalem]{ulem}
\usepackage[title]{appendix}
\usepackage{comment}
\usepackage{float}
\usepackage{mathtools}

\usepackage{ subfigure }

\usepackage{hyperref}

\usepackage[utf8]{inputenc}
\usepackage{amsthm}

\usepackage[T1]{fontenc}
\usepackage[utf8]{inputenc}
\usepackage{babel}

\newcommand\bigzero{\makebox(0,0){\text{\huge0}}}
\newcolumntype{C}[1]{>{\centering\arraybackslash$}m{#1}<{$}}
\newlength{\mycolwd} 
\begin{document}

\title{Group synchrony, parameter mismatches, and intragroup connections}

\author{Shirin Panahi} 
\author{Francesco Sorrentino}
\thanks{fsorrent@unm.edu}
\address{University of New Mexico, Albuquerque New Mexico 87106, USA}

\begin{abstract}
Group synchronization arises when two or more synchronization patterns coexist in a network formed of oscillators of different types, with the systems in each group synchronizing on the same time-evolution, but systems in different groups synchronizing on distinct time-evolutions. 
Group synchronization has been observed and characterized when the systems in each group are identical and the couplings between the systems satisfy specific conditions. By relaxing these constraints and allowing them to be satisfied in an approximate rather than exact way, we observe that stable group synchronization may still occur in the presence of small deviations of the parameters of the individual systems and of the couplings from their nominal values. We analyze this case and provide necessary and sufficient conditions for stability through a master stability function approach, {which also allows us to quantify the synchronization error}. We also investigate the stability of group synchronization in the presence of intra group connections and for this case, extend some of the existing results in the literature. Our analysis points out a broader class of matrices describing intra group connections for which the stability problem can be reduced in a low-dimensional form.
\end{abstract}

\maketitle

\section{Introduction}
{A large literature has considered synchronization in networks of coupled oscillators, see for example the review in Ref. \cite{SReport}. In the case of identical oscillators, stability of the synchronous solution for arbitrary network topologies can be investigated through the master stability function approach, introduced in Ref. \cite{Pe:Ca}. More complex patterns of synchronous dynamics may arise in networks formed of different types of oscillators.}
Group synchronization in networks of coupled oscillators was first analyzed in \cite{NSG}, {followed by further analysis in Ref. \cite{clustergroup} and an experimental realization of group synchronization \cite{williams2013experimental}.} {Moreover, group synchronization in networks of fractional-order chaotic oscillators was investigated in Ref. \cite{yang2015group}}.

According to the definition in Ref. \cite{NSG}, a \emph{group} is a set of oscillators characterized by the same (uncoupled) dynamics, with oscillators of different types belonging to different groups. {Group synchronization} is achieved when the oscillators in each group synchronize on the same time evolution, with these synchronized dynamics being different from group to group. Cluster synchronization can be considered as a particular case of group synchronization, but for which the oscillators in different clusters are of the same type. {Most recently, a canonical transformation for simultaneous block diagonalization of matrices has been proposed to decouple the cluster synchronization stability problem into subproblems of minimal dimensionality \cite{panahi2021cluster}.} {The more general case of cluster synchronization in multilayer networks formed of oscillators of different types has been studied in Ref. \cite{della2020symmetries}.}

An experimental realization of group synchronization was performed in Ref. \cite{williams2013experimental}, where stable group synchronization was observed under different configurations for a small network of four coupled optoelectronic oscillators. This observation suggests that group synchronization is robust with respect to small parametric mismatches that are inevitable in experiments. The analysis of stability of the group-synchronous solution in the presence of such small parametric mismatches (affecting both the individual nodes being coupled and the strengths of the couplings between them) is studied in this paper. The effects of parameter mismatches in a network of \emph{identical} oscillators achieving complete synchronization have been studied in Refs. \cite{restr_bubbl,Su:Bo:Ni,SOPO} { and achieving cluster synchronization in Refs. \cite{sorrentino2016approximate,cho2019concurrent}. A second order expansion to study complete synchronization in networks of identical nodes with parameter mismatches has been studied in Refs. \cite{acharyya2012synchronization,acharyya2015synchronization}. However, to the best of our knowledge, no paper has investigated the role of parameter mismatches in networks formed of systems of different types, which is the main focus of this paper.}  The other subject of this paper is the study of the conditions under which group synchronization may arise in the presence of intragroup connections, which was first considered in Ref. \cite{clustergroup}.

\section{Group synchronization in the presence of parametric mismatches}\label{II}

Though our results can be generalized to the case of an arbitrary number of groups (for the case in which the stability of the group-synchronous solution can be reduced in a low-dimensional form \cite{clustergroup}), for simplicity in what follows we will focus on the case of two groups. For the time being, we also maintain the assumption introduced in \cite{NSG} that the network topology is bipartite, i.e., connections exist only from group $X$ to group $Y$ and vice versa (this assumption will be removed later in Sec.\ref{III}).

We first present the ideal case that parameter mismatches are absent. For this very special case, the dynamical equations are:

\begin{subequations}\label{nom}
	\begin{equation}\label{noma}
		\dot{\bx}_{i} = \bF \Bigl( \bx_{i}, u_i^x \Bigr)= \bF \Bigl(\bx_i, \sum_{j=1}^{N_{y}} A_{ij}^{\mathrm{NOM}} \bH(\by_{j}) \Bigr), \quad i= 1,\cdots, N_{x}.
		\end{equation}
		\begin{equation}\label{nomb}
		\dot{\by}_{j} = \bG \Bigl( \by_{j}, u_j^y \Bigr)= \bG\Bigl(\by_j, \sum_{i=1}^{N_{x}} B_{ji}^{\mathrm{NOM}} \bL(\bx_{i}) \Bigr), \quad j= 1,\cdots, N_{y}.
	\end{equation}
\end{subequations}
where $\bx_i$ ($\by_j$) is an $n_x$-dimensional ($n_y$-dimensional) state vector of systems in group $\bX$ and in group $\bY$ respectively, $\bu_i^x=\sum_{j=1}^{N_{y}} A_{ij}^{\mathrm{NOM}} \bH(y_{j})$ $[\bu_j^y=\sum_{i=1}^{N_{x}} B_{ji}^{\mathrm{NOM}} \bL(x_{i})]$ is the input received by node $i$ in group $\bX$ ($j$ in group $\bY$). The functions  $\bF: \mathbf{R}^{n_x} \times \mathbf{R}^{n_x} \rightarrow  \mathbf{R}^{n_x}$ and $\bG: \mathbf{R}^{n_y} \times \mathbf{R}^{n_y} \rightarrow  \mathbf{R}^{n_y}$ define the time evolution of the systems in group $\bX$ and group $\bY$ respectively, and  the interaction  functions $\bH: \mathbf{R}^{n_y} \rightarrow  \mathbf{R}^{n_x}$ and $\bL: \mathbf{R}^{n_x} \rightarrow  \mathbf{R}^{n_y}$ define the output of systems in group $\bY$ and group $\bX$, respectively. $N_x$ is the number of systems in group $\bX$ and $N_y$ is the number of systems in group $\bY$. In what follows we assume without loss of generality that $N_x\geq N_y$. The input $\bu_i^x$ received by node $i$ in group $X$ ($\bu_j^y$ received by node $j$ in group $\bY$) corresponds to a superposition of the outputs from the nodes in group $\bY$ (group $\bX$) through the coefficients $A_{ij}^{NOM}$ ($B_{ji}^{NOM}$). $A^{\mathrm{NOM}}$ is an $N_{x} \times N_{y}$ coupling matrix, whose entries ${A_{ij}^{\mathrm{NOM}}}$ represent the
nominal strength of the direct interaction form system $j$ in the $\bY$-group to system $i$ in the $\bX$-group. Analogously,
$B_{ji}^{\mathrm{NOM}}$ is an $N_{y} \times N_{x}$ coupling matrix, whose entries $B_{ji}^{\mathrm{NOM}}$ represent the
nominal strength of the direct interaction from system $i$ in the $\bX$-group to $j$ in the $\bY$-group. Note that Eqs. \eqref{nom} are a generalization of those considered in Refs. \cite{NSG,clustergroup}.

In what follows, we consider the following two types of small parameter mismatches from nominal conditions: (i) deviations from identicality of the uncoupled dynamics of the individual systems in each group and (ii) deviations from nominal conditions of the coupling strengths between systems.

If the following conditions hold,
\begin{subequations}
	\label{eq:condition1}
	\begin{align}
		 \sum_{j=1}^{N_{y}} A_{ij}^{\mathrm{NOM}} &= a \neq 0, \quad i= 1,\cdots, N_{x}. \\
		 \sum_{i=1}^{N_{x}} B_{ji}^{\mathrm{NOM}} &= b \neq 0, \quad j= 1,\cdots, N_{y}.
	\end{align}
\end{subequations}
then the synchronous solution exists,
\begin{subequations} \label{equs}
	\begin{align}
		\dot{\bx}_{s} &= \bF\Bigl(\bx_{s}, a \bH(\by_{s}) \Bigr), \\
		\dot{\by}_{s} &= \bG\Bigl(\by_{s}, b \bL(\bx_{s}) \Bigr).
	\end{align}
\end{subequations}
We note here that by replacing $a \bH \rightarrow \bH $ and $A^{NOM}/a \rightarrow A^{NOM}$ ($b \bL \rightarrow \bL$ and $B^{NOM}/b \rightarrow B^{NOM}$), it is always possible to set $a=b=1$. {Hence, without loss of generality, in what follows we set $a=b=1$. This has the significant advantage that the synchronous solution, which obeys Eqs.\ \eqref{equs} with $a=b=1$, is independent of the particular choice of the bipartite network.}

Stability of this synchronized solution can be reduced in a master stability form \cite{NSG}. Unfortunately, exact
satisfaction of condition \eqref{eq:condition1} is difficult to implement in experiments and in real-world situations. The other requirement whose exact satisfaction is hardly accomplished in experiments is that the evolution of all the systems in the $\bX$-group (in the $\bY$-group) is exactly characterized by the same function $F$ ($G$).
What can be realistically achieved is that $\bF(\bx_i,\bu_i^x) \rightarrow \bF(\bx_i,\bu_i^x,\mu_i)$, $i=1,...,N_x$ in Eq.\eqref{noma} and $\bG(\by_j,\bu_j^y) \rightarrow \bG(\by_j,\bu_j^y,{\nu_j})$, $j=1,...,N_y$ in Eq. \eqref{nomb}
where $\mu_i$ and $\nu_j$ are scalar parameters that slightly vary from oscillator to oscillator \footnote{Even though we do not consider the case of two or more parameters that vary from system to system, the extension of our results to this case is straightforward.}. Analogously, it is realistic to assume that the couplings $A_{ij}$ and $B_{ji}$ are affected by small mismatches with respect to their nominal values,
\begin{subequations}
	\label{eq:AandB}
	\begin{align}
		A_{ij} &= A_{ij}^{\mathrm{NOM}} + \delta A_{ij} , \\
		B_{ji} &= B_{ji}^{\mathrm{NOM}} + \delta B_{ji} ,
	\end{align}
\end{subequations}
where $\delta A_{ij}$ and $ \delta B_{ji}$ are small deviations. 
Under these assumptions, Eqs. \eqref{nom} become
\begin{subequations}
	\label{eq:new_equations}
	\begin{align}
		\dot{\bx}_{i} &= \bF \Bigl(\bx_{i}, \sum_{j=1}^{N_{y}} A_{ij} \bH(\by_{j}), \mu_i \Bigr), \quad i= 1,\cdots, N_{x}, \\
		\dot{\by}_{j} &= \bG \Bigl(\by_{j}, \sum_{i=1}^{N_{x}} B_{ji} \bL(\bx_{i}), \nu_j \Bigr), \quad j= 1,\cdots, N_{y},
	\end{align}
\end{subequations}
which describe the dynamics of the two groups under realistic circumstances, such as those that would be observed in any experimental setting.

We write $\mu_i=\bar{\mu}+\delta \mu_i$ where $\bar{\mu}= (N_x)^{-1} \sum_{i=1}^{N_x} \mu_i$  and  $\delta \mu_i$ is a small deviation. Similarly, we write $\nu_j=\bar{\nu}+\delta \nu_j$ where $\bar{\nu}= (N_y)^{-1} \sum_{j=1}^{N_y} \nu_j$ and $\delta \nu_j$ is a small deviation.
Note that by construction $\sum_{i=1}^{N_x} \delta \mu_i=0$ and $\sum_{j=1}^{N_y} \delta \nu_j=0$.

We can also write 
\begin{subequations}
	\label{eq:condition2}
	\begin{align}
		 \sum_{j=1}^{N_{y}} A_{ij} &= a +   \sum_{j=1}^{N_{y}} \delta A_{ij}= a + \delta \bar{a}  + \delta a_{i},  \\
		 \sum_{i=1}^{N_{x}} B_{ji} &= b +   \sum_{i=1}^{N_{x}} \delta B_{ji}= b + \delta \bar{b}  + \delta b_{j},
	\end{align}
\end{subequations}
where
\begin{subequations}
	\label{eq:condition3}
	\begin{align}
		 \delta \bar{a} &=  N_{x}^{-1} \sum_{i=1}^{N_{x}} \sum_{j=1}^{N_{y}}\delta A_{ij} ,\\
		 \delta \bar{b} &=  N_{y}^{-1} \sum_{i=1}^{N_{x}} \sum_{j=1}^{N_{y}}\delta B_{ji} ,
	\end{align}
\end{subequations}
are the average sums of the rows of the matrices $A$ and $B$ and

\begin{subequations}
	\label{eq:condition4}
	\begin{align}
		 \delta a_{i}  &= \Bigg( \sum_{j=1}^{N_{y}} \delta A_{ij}  \Bigg) -\delta \bar{a} = \Bigg( \sum_{j=1}^{N_{y}} \delta A_{ij}  \Bigg) - N_{x}^{-1} \sum_{i=1}^{N_{x}} \sum_{j=1}^{N_{y}}\delta A_{ij} ,\\
		 \delta b_{j} &=  \Bigg( \sum_{i=1}^{N_{x}} \delta B_{ji}  \Bigg) -\delta \bar{b} =\Bigg( \sum_{i=1}^{N_{x}} \delta B_{ji}  \Bigg) - N_{y}^{-1} \sum_{i=1}^{N_{x}} \sum_{j=1}^{N_{y}}\delta B_{ji} ,
	\end{align}
\end{subequations}
are small deviations. The deviations $\delta a_{i}$ and $\delta b_{j}$ are calculated with respect to the average row-sums
$\delta \bar{a}$ and $\delta \bar{b}$, hence they sum to zero, that is, $\sum_{i=1}^{N_{x}} \delta a_{i}=0$ and
$\sum_{j=1}^{N_{y}} \delta b_{j}=0$.

Unfortunately, for the case of equations \eqref{eq:new_equations}, different from the case of equations \eqref{nom} an exact synchronous solution does not exist. It is possible, however, that the individual trajectories stabilize in a nearly synchronous state, where the trajectories of the nodes in the $\bX$-group (and in the $\bY$-group) remain close to each other, i.e., for which both
\begin{subequations} \label{var}
 \begin{align}
  \delta \bx_{i}(t) &= (\bx_{i}(t) - \bar{\bx}(t)), \\
  \delta \by_{j}(t) &= (\by_{j}(t) - \bar{\by}(t)),
 \end{align}
\end{subequations}
remain small in time, where $\bar{\bx}(t)$ and $\bar{\by}(t)$ are the average solutions,

\begin{subequations}
	\label{eq:condition5}
	\begin{align}
		 \bar{\bx}(t) &= (N_{x}^{-1})\sum_{i=1}^{N_{x}} \bx_{i}(t), \\
		 \bar{\by}(t) &= (N_{y}^{-1})\sum_{j=1}^{N_{y}} \by_{j}(t), 		
	\end{align}
\end{subequations}
obeying,
\begin{subequations}
 \label{eq:dynamicsCondition}
    \begin{align}
        \dot{\bar{\bx}} &= (N_{x})^{-1} \sum_{i=1}^{N_{x}} \bF \Bigl(\bx_{i}, \sum_{j=1}^{N_{y}} A_{ij} \bH(y_{j}), \mu_i \Bigr),  \\
        \dot{\bar{\by}} &= (N_{y})^{-1} \sum_{j=1}^{N_{y}} \bG \Bigl( \by_{j}, \sum_{i=1}^{N_{x}} B_{ji} \bL(\bx_{i}), \nu_j \Bigr).
    \end{align}
\end{subequations}

By differentiating \eqref{var} with respect to time, we obtain
\begin{subequations}
 \label{vardif}
    \begin{align}
       \delta \dot{\bx}_i &=\Bigl[ \bF \Bigl(\bx_{i}, \sum_{j=1}^{N_{y}} A_{ij} \bH(\by_{j}), \mu_i \Bigr)-(N_{x})^{-1} \sum_{i=1}^{N_{x}} \bF \Bigl(\bx_{i}, \sum_{j=1}^{N_{y}} A_{ij} \bH(\by_{j}), \mu_i \Bigr) \Bigr], \quad i= 1,\cdots, N_{x},  \\
        \delta \dot{\by}_j &= \Bigl[ \bG \Bigl(\by_{j}, \sum_{i=1}^{N_{x}} B_{ji} \bL(\bx_{i}), \nu_j \Bigr)- (N_{y})^{-1} \sum_{j=1}^{N_{y}} \bG \Bigl( \by_{j}, \sum_{i=1}^{N_{x}} B_{ji} \bL(\bx_{i}), \nu_j \Bigr) \Bigr], \quad j= 1,\cdots, N_{y}.
    \end{align}
\end{subequations}

By expanding to first order the function $\bF$ in (\ref{vardif}a) about the point $(\bar{\bx},\bH(\bar{\by}),\bar{\mu})$ and the function $\bG$ in (\ref{vardif}b) about the point $(\bar{\by},\bL(\bar{\bx}),\bar{\nu})$, we obtain,

 \begin{align*}
\bF \Bigl(\bx_{i}, \sum_{j=1}^{N_{y}} A_{ij} \bH(\by_{j}), \mu_i \Bigr)  & \simeq \bF \Bigl(\bar{\bx}, \sum_{j=1}^{N_{y}} A_{ij}^{\mathrm{NOM}} \bH(\bar{\by}), \bar{\mu} \Bigr)+ \mathrm{D}\bF_x \Bigl(\bar{\bx}, \sum_{j=1}^{N_{y}} A_{ij}^{\mathrm{NOM}} \bH(\bar{\by}), \bar{\mu} \Bigr) \delta {\bx}_i \\  &+ \mathrm{D}\bF_u \Bigl(\bar{\bx}, \sum_{j=1}^{N_{y}} A_{ij}^{\mathrm{NOM}} \bH(\bar{\by}), \bar{\mu} \Bigr) \Bigl[\sum_{j=1}^{N_{y}}  A_{ij}^{\mathrm{NOM}} \mathrm{D}\bH(\bar{\by}) \delta \by_j + \sum_{j=1}^{N_{y}} \bH(\bar{\by}) \delta A_{ij}  \Bigr]  \\ &+ \mathrm{D}F_\mu \Bigl(\bar{\bx}, \sum_{j=1}^{N_{y}} A_{ij}^{\mathrm{NOM}} \bH(\bar{\by}), \bar{\mu} \Bigr) \delta \mu_i\\
 \bG \Bigl(\by_{j}, \sum_{i=1}^{N_{x}} B_{ji} \bL(\bx_{i}), \nu_j \Bigr)  & \simeq \bG \Bigl(\bar{\by}, \sum_{i=1}^{N_{x}} B_{ji}^{\mathrm{NOM}} \bL(\bar{\bx}), \bar{\nu} \Bigr)+ \mathrm{D}\bG_y \Bigl(\bar{\by}, \sum_{i=1}^{N_{x}} B_{ji}^{\mathrm{NOM}} \bL(\bar{\bx}), \bar{\nu} \Bigr) \delta {\by}_j \\  &+ \mathrm{D}\bG_u \Bigl(\bar{\by}, \sum_{i=1}^{N_{x}} B_{ji}^{\mathrm{NOM}} \bL(\bar{\bx}), \bar{\nu} \Bigr) \Bigl[\sum_{i=1}^{N_{x}}  B_{ji}^{\mathrm{NOM}} \mathrm{D}\bL(\bar{\bx}) \delta \bx_i + \sum_{i=1}^{N_{x}} \bL(\bar{\bx}) \delta B_{ji}  \Bigr]  \\ &+ \mathrm{D}\bG_\nu \Bigl(\bar{\by}, \sum_{i=1}^{N_{x}} B_{ji}^{\mathrm{NOM}} \bL(\bar{\bx}), \bar{\nu} \Bigr) \delta \nu_i\\
\end{align*}
Then Eqs.\ \eqref{vardif} can be rewritten,

\begin{subequations}
\label{eq:SystemOfEqs}
 \begin{align}
    \delta \dot{\bx}_{i}(t) &= \mathrm{D} \bF_x^* \delta \bx_{i}(t) + \sum_{j=1}^{N_{y}} \tilde{A}_{ij} \mathrm{D} \bF_u^* \mathrm{D} \bH_y(\bar{\by}) \delta \by_{j} + \bU^x_i, \quad i=1,...,N_x, \\
    \delta \dot{\by}_{j}(t) &= \mathrm{D} \bG_y^* \delta \by_{j}(t) + \sum_{i=1}^{N_{x}} \tilde{B}_{ji} \mathrm{D} \bG_u^* \mathrm{D} \bL_x(\bar{\bx}) \delta \bx_{i} + \bU^y_j, \quad j=1,...,N_y,
 \end{align}
\end{subequations}
where $\bU^x_i=\mathrm{D} \bF_u^* \bH(\bar{\by}) \delta a_i+ \mathrm{D} \bF_\mu^* \delta \mu_i$ and $\bU^y_j=\mathrm{D} \bG_u^* \bL(\bar{\bx})\delta b_j+ \mathrm{D} \bG_\nu^* \delta \nu_j$, 
and the superscript $^*$ denotes that the partial derivatives of the functions $\bF$ and $\bG$ are evaluated about $(\bar{\bx},\bH(\bar{\by}),\bar{\mu})$ and $(\bar{\by},\bL(\bar{\bx}),\bar{\nu})$, respectively. The matrices $\tilde{A}=\{ \tilde{A}_{ij} \}$ and
$\tilde{B}=\{ \tilde{B}_{ij} \}$ in \eqref{eq:SystemOfEqs} are defined as follows,
\begin{subequations} \label{TILDE}
 \begin{align}
     \tilde{A}_{ij}  & = (A_{ij}^{NOM}-a_{j}), \qquad i=1,\dots,N_{x}, \quad j=1,\dots, N_{y}, \\
     \tilde{B}_{ji}  & = (B_{ji}^{NOM}-b_{i}), \qquad i=1,\dots,N_{x}, \quad j=1,\dots, N_{y},
 \end{align}
\end{subequations}
where $a_{j} = (N_{x})^{-1} \sum_{i=1}^{N_{x}} A_{ij}^{NOM}$ is the mean value over the elements in the $j^{th}$ column of the matrix $A^{NOM}$
and $b_{i} = (N_{y})^{-1} \sum_{j=1}^{N_{y}} B_{ji}^{NOM}$ is the mean value over the elements in the $i^{th}$ column of the matrix $B^{NOM}$. Note that in order to obtain \eqref{eq:SystemOfEqs} we have used Eqs. \eqref{eq:condition4} and the following properties $\sum_{i=1}^{N_{x}} \delta \mu_{i}=0$, $\sum_{j=1}^{N_{y}} \delta \nu_{j}=0$, $\sum_{i=1}^{N_{x}} \delta a_{i}=0$, $\sum_{j=1}^{N_{y}} \delta b_{j}=0$,
$\sum_{i=1}^{N_{x}} \delta \bx_{i}=0$ and $\sum_{j=1}^{N_{y}} \delta \by_{j}=0$. {Note also that the two matrices $\tilde{A}$ and $\tilde{B}$ have sums over both their columns and rows equal to zero.}

In what follows, we will study stability of the system \eqref{eq:SystemOfEqs}. If this system is found to be stable, that implies that the trajectories \eqref{nom} will tend to remain close to the average solution \eqref{eq:condition5} [see the definition of the variations in \eqref{var}].

{In what follows, we assume that for each connection from node $i$ in group $\bX$ to node $j$ in group $\bY$, there is a connection from node $j$ in group $\bY$ to node $i$ in group $\bX$ and vice versa. We note that we must still satisfy the assumption (which we imposed at the beginning of the paper without loss of generality) that the sums over the entries in the rows of both matrices $A^{NOM}$ and $B^{NOM}$ are equal to one. With these considerations in mind, we take
$B^{NOM}= e {A^{NOM}}^{T}$, where $e=N_y/N_x$. It then follows $\tilde{B}=e \tilde{A}^T$.}

We then rewrite Eq.\ \eqref{eq:SystemOfEqs} in vectorial form,
\begin{subequations}
\label{vec:SystemOfEqs}
 \begin{align}
 \delta \dot{\bx}=[I_{N_x} \otimes \mathrm{D} \bF_x^*]\delta \bx + [\tilde{A} \otimes \mathrm{D} \bF_u^* \mathrm{D} \bH_y(\bar{\by})]\delta \by +\bU_x\\
  \delta \dot{\by}=[I_{N_y} \otimes \mathrm{D} \bG_y^*]\delta \by+ [e \tilde{A}^T \otimes \mathrm{D} \bG_u^* \mathrm{D} \bL_x(\bar{\bx})]\delta \bx +\bU_y
 \end{align}
\end{subequations}

By the substitution $\delta \by \leftarrow \delta \by/\sqrt{e}$, Eq.\ \eqref{vec:SystemOfEqs} becomes,
\begin{subequations}
\label{vec:SystemOfEqs2}
 \begin{align}
 \delta \dot{\bx}=[I_{N_x} \otimes \mathrm{D} \bF_x^*]\delta \bx + [\sqrt{e}\tilde{A} \otimes \mathrm{D} \bF_u^* \mathrm{D} \bH_y(\bar{\by})]\delta \by +\bU_x\\
  \delta \dot{\by}=[I_{N_y} \otimes \mathrm{D} \bG_y^*]\delta \by+ [\sqrt{e} \tilde{A}^T \otimes \mathrm{D} \bG_u^* \mathrm{D} \bL_x(\bar{\bx})]\delta \bx +\bU_y/\sqrt{e}
 \end{align}
\end{subequations}

By stacking all perturbation vectors together in one vector ${\bz}=[\delta {\bx}^T,\delta {\by}^T]^T$ and by introducing the vector $\bU=[\bU_x^T,\bU_y^T/\sqrt{e}]^T$, Eq.\ \eqref{vec:SystemOfEqs} can be rewritten as
    \begin{equation} \label{z}
    \begin{array}{c}
\dot{\bz}=
\begin{pmatrix}
  I_{N_x} \otimes \mathrm{D} \bF_x^* & 0\\ 
  0 & {0}_{N_y} \otimes \mathrm{D} \bG_y^*
    \end{pmatrix}\bz+
    \begin{pmatrix}
  0_{N_x} \otimes \mathrm{D} \bF_x^* & 0\\ 
  0 & I_{N_y} \otimes \mathrm{D} \bG_y^*
    \end{pmatrix}\bz+\\\\
   \sqrt{e} \begin{pmatrix}
  0 &  \tilde{A} \otimes \mathrm{D} \bF_u^* \mathrm{D} \bH_y(\bar{\by})\\ 
    \tilde{A}^T \otimes \mathrm{D} \bG_u^* \mathrm{D} \bL_x(\bar{\bx}) & 0
    \end{pmatrix}\bz+\bU  

    \end{array}
\end{equation}
We define the matrices,
\begin{equation}\label{E_i}
    E^x=\begin{pmatrix}
  I_{N_x} & 0\\ 
  0 & {0}_{N_y}
    \end{pmatrix}
    \quad
        E^y=\begin{pmatrix}
  0_{N_x} & 0\\ 
  0 & {I}_{N_y}
    \end{pmatrix}
\quad
   \tilde{M}= \begin{pmatrix}
  0_{n_x} & \tilde{A}\\ 
   \tilde{A}^T & {0}_{n_y}
    \end{pmatrix}
\end{equation}
and seek to compute the $N_x+N_y$-dimensional transformation matrix $T$ that simultaneously block diagonalizes the matrices in Eq.\ \eqref{E_i}, 
\begin{equation}
    T=\mathcal{SBD}(E^x,E^y,\tilde{M}),
\end{equation}
where here by $\mathcal{SBD}$ we indicate that the transformed matrices $T E^x T^T$, $T E^y T^T$, and $T \tilde{M} T^T$  are in the same block-diagonal form with blocks of minimal dimension \cite{SBD1,SBD2,SBD3,Ir:SO,zhang2020symmetry}.
To compute the matrix $T$ we first need to find a matrix $P$ that commutes with the symmetric matrices $E^x$, $E^y$, and $\tilde{M}$. The matrix $T$ has for columns the eigenvectors of the matrix $P$ \cite{SBD2}.
In order for the matrix $P$ to commute with the diagonal matrices $E^x$ and $E^y$,  it needs to be in the following block-diagonal form,
\begin{equation}\label{Pblock}
P=\begin{pmatrix} 
  P_{1} & 0 \\
  0 & P_{2} 
\end{pmatrix}
\end{equation}
where the square block $P_{1}$ has dimension $N_{x}$  and the square block $P_2$ has dimension $N_y$.
Finally, the matrix $P$ in \eqref{Pblock} needs also to commute with $\tilde{M}$.
Then the following two relations must be satisfied,
\begin{subequations}
    \begin{align}
        \tilde{A} P_2= & P_1 \tilde{A} \\
        \tilde{A}^T P_1= & P_2 \tilde{A}^T,
    \end{align}
\end{subequations}
 A simple solution to the above set of equations is given by
 \begin{subequations}
     \begin{align}
         P_1= & \tilde{A} \tilde{A}^T \\
         P_2= & \tilde{A}^T \tilde{A}.
     \end{align}
 \end{subequations}
 From Eq.\ \eqref{Pblock} it follows that the matrix $T$ has the following block-diagonal structure,
\begin{equation}
\label{Tblock}
T=\begin{pmatrix} 
  T^{x} & 0 \\
  0 & T^{y}, 
\end{pmatrix}
\end{equation}
where the columns of the matrix $T^x$ are the eigenvectors of the matrix $\tilde{A}\tilde{A}^T$
and the columns of the matrix $T^y$ are the eigenvectors of the matrix $\tilde{A}^T\tilde{A}$. 
As both matrices $\tilde{A}\tilde{A}^T$ and $\tilde{A}^T\tilde{A}$ are symmetric, the matrices $T^x$ and $T^y$ are orthogonal. Hence, $T$ is also an orthogonal matrix.

We can then define the $N_x n_x+N_y n_y$-dimensional orthogonal matrix 
\begin{equation}
\hat{T}=(T^x \otimes I_{n_x}) \oplus (T^y \otimes I_{n_y})
\end{equation}
and use it to block-diagonalize Eq.\ \eqref{z}. We introduce $\bw=\hat{T} \bz$, then we have:
 \begin{equation}\label{w}
\dot{\bw}=\begin{pmatrix}
  I_{N_x} \otimes \mathrm{D} \bF_x^* & 0\\ 
  0 & I_{N_y} \otimes \mathrm{D} \bG_y^*
    \end{pmatrix} \bw + \sqrt{e}
    \begin{pmatrix}
  0 & S_1 \otimes \mathrm{D} \bF_u^* \mathrm{D} \bH_y(\bar{\by})\\ 
   S_{1}^T \otimes \mathrm{D} \bG_u^* \mathrm{D} \bL_x(\bar{\bx}) & 0
    \end{pmatrix} \bw+T \bU
\end{equation}
where 
\begin{equation} \label{S1}
S_1={T^x} \tilde{A} {T^y}^T.
\end{equation}
As  mentioned before, both $T^x$ and $T^y$ are orthogonal matrices, hence ${T^x}^{T}{T^x}=I_{N_x}$ and ${T^y}^{T}T^y=I_{N_y}$. From \eqref{S1}, we see that $S_1$ is the $N_x$ by $N_y$ matrix whose entries ${S_1}_{kk}=s_{\kappa}$ are equal to the singular values of the matrix $\tilde{A}$ (all the other entries are zero).

{
There are many possible numerical methods to compute the singular values of a matrix \cite{golub1965calculating}.
We note here that the eigenvalues of the matrix $\tilde{M}$ in \eqref{E_i} are equal to $\{ \pm s_1, \pm s_2,..., \pm s_{r}\} \cup \{\underbrace{0,0,...,0}_\text{$N_x-N_y$ times} \}$ \cite{NSG}. From a computational point of view the method of computing the singular values from the eigenvalues of the matrix $\tilde{M}$ is more stable than direct calculation of the matrix $S_1$ in \eqref{S1}.}

We note that through the transformation $\bw=\hat{T} \bz$, each row of the matrix $T^x$, say $T^{x\kappa}$, is associated with either a singular value $s_\kappa$ or a zero value; analogously, each row of the matrix $T^y$, say $T^{y\kappa}$, is associated with  a singular value $s_\kappa$.

We then introduce the matrix:
\begin{equation}
Q= \left( \begin{array}{cc} 
     0 & S_1 \\
   S_1^T  & 0
    \end{array}\right)=
   \left( \begin{array}{ccccccc|cccc}
  & & & & & & & s_1 & 0 & \cdots & 0\\
  & & & & & & & 0 & s_2 & \cdots & 0\\
   & & & & & & & 0 & 0 & \ddots & 0\\
   & & & \bigzero & & & & 0 & 0 & \cdots & s_r\\
   & & & & & & & 0 & 0 & \cdots & 0\\
   & & & & & & & \vdots & \vdots & \ddots & \vdots\\
   & & & & & & & 0 & 0 & \cdots & 0\\
\hline
 s_1 & 0 & 0 & 0 & 0 & \cdots & 0 & & & & \\
0 &  s_2 & 0 & 0 & 0 & \cdots & 0 & &  & & \\
\vdots & \vdots & \ddots & \vdots & \vdots & \ddots & \vdots & &  & \bigzero & \\
0 & 0 & 0 &  s_r & 0 & \cdots & 0 & & & & \\
\end{array}\right)
\end{equation}
Based on our assumption that $N_x \geq N_y$, we have that $r=N_y$. We call $d=(N_x-N_y)$, $d+r=N_x$.

Now by permuting rows and columns of matrix $Q$, we obtain the block diagonal matrix,
\begin{equation} \label{Q}
    \tilde{Q}=\left(
    \begin{array}{ccccc|ccc}
      \tilde{Q}_1 & 0 & 0 & \cdots & 0 & 0 & \cdots & 0\\
      0 & \tilde{Q}_2 & 0 & \cdots & 0 & 0 & \cdots & 0\\
      \vdots & \vdots & \vdots & \ddots & \vdots & \vdots & \ddots & \vdots\\
      0 & 0 & 0& \cdots & \tilde{Q}_r & 0 & \cdots & 0\\
      \hline
      0 & 0 & 0& \cdots & 0 & 0 & \cdots & 0\\
      \vdots & \vdots & \vdots & \ddots & \vdots & \vdots & \ddots & \vdots\\
      0 & 0 & 0& \cdots & 0 & 0 & \cdots & 0\\
\end{array}\right)
\end{equation}
where each $\tilde{Q}_i \quad i=1,2,\cdots, r$ is a $2$ by $2$ block,
\begin{equation}
    \tilde{Q}_\kappa=\begin{pmatrix}
      0 & s_\kappa\\
      s_\kappa & 0
    \end{pmatrix}
    \quad
    \kappa=1, 2, \cdots, r.
\end{equation}
One of the blocks $\kappa=1$ corresponds to motion parallel to synchronization manifold and the other $r-1$ blocks, $i=2,...,r$ correspond to motion transverse to the synchronization manifold \cite{NSG}. By applying the same permutation to the first term on the right hand side of Eq. \eqref{w}, for each $\kappa=1,..,r$ we can write:
\begin{equation} \label{wj}
\dot{\bp}_{\kappa}=\Biggl[\begin{pmatrix}
    \mathrm{D} \bF_x^* & 0\\
    0 &   \mathrm{D} \bG_y^*
    \end{pmatrix}+\begin{pmatrix}
      0 & \lambda_\kappa \mathrm{D} \bF_u^* \mathrm{D} \bH_y(\bar{\by}) \\
          \lambda_\kappa \mathrm{D}  \bG_u^* \mathrm{D} \bL_x(\bar{\bx})& 0
    \end{pmatrix}\Biggr]\bp_{\kappa} + \Biggl[ \begin{matrix}
       {\Big( \sum_{i}^{N_x} {T_i}^{x \kappa} \delta a_{i} \Big)} \mathrm{D} \bF_u^* H(\bar{y}) + {\Big( \sum_{i}^{N_x} {T_i}^{x \kappa} \delta \mu_{i} \Big)}  \mathrm{D} \bF_\mu^*\\
       {\Big( \frac{\sum_{i}^{N_y} {T_i}^{y \kappa} \delta b_{i}}{\sqrt{e}} \Big)} \mathrm{D} \bG_u^* L(\bar{x}) + {\Big( \frac{\sum_{i}^{N_y} {T_i}^{y \kappa} \delta \nu_{i}}{\sqrt{e}} \Big)}  \mathrm{D} \bG_\nu^*
    \end{matrix} \Biggl],
\end{equation}
where $\lambda_\kappa=\sqrt{e} s_\kappa $, $T^{x\kappa}_{i}$ ($T^{x\kappa}_{i}$) is entry $i$ of row $T^{x\kappa}$ ($T^{x\kappa}$).  
Analogously, to the right-lower block of the matrix $\tilde{Q}$ (Eq.\ \eqref{Q}), we can associate $d$ equations,
\begin{equation} \label{hj}
\dot{\bq}_{\kappa}= \mathrm{D} \bF_x^*
   \bq_{\kappa} + {\Big( \sum_{i} T^{x\kappa}_{i} \delta a_{i} \Big)} \mathrm{D} \bF_u^* H(\bar{y}) + {\Big( \sum_{i} T^{x\kappa}_{i} \delta \mu_{i} \Big)}  \mathrm{D} \bF_\mu^*,
    \quad
    \kappa=1,2,\cdots, d.
\end{equation}

The homogeneous part of Eqs.\ \eqref{wj} and \eqref{hj} coincides with the reduction obtained in Ref.\ \cite{NSG}.

The vector $\bw$ can be written as $[{\bp}_1^T,{\bp}_2^T,...,{\bp}_r^T,\bq_1^T,\bq_2^T,...,\bq_p^T]^T$, where ${\bp}_1^T$ is a parallel perturbation and all the remaining perturbations are transverse.

We now define the following master stability functions associated with Eqs.\ \eqref{wj} and \eqref{hj},
\begin{subequations}\label{eq:MSFn}
  \begin{equation}\label{eq:MSFna}
     \mathcal{M}_\kappa(s_\kappa,\alpha_\kappa,\beta_\kappa,\gamma_k,\delta_\kappa)  =  \lim_{\tau \to \infty} \sqrt{ \tau^{-1} \int_{0}^{\tau} \| {\bp}_\kappa(t) \|^{2} dt},
     \end{equation}
     \begin{equation}\label{eq:MSFnb}
     \mathcal{M}_{0}(0,\alpha_\kappa,\beta_\kappa)  =  \lim_{\tau \to \infty} \sqrt{ \tau^{-1} \int_{0}^{\tau} \| \bq_\kappa(t) \|^{2} dt},
  \end{equation}
\end{subequations}
where $\alpha_\kappa={( \sum_{i} T^{x \kappa}_{i} \delta a_{i} )}$, $\beta_\kappa={( \sum_{i} T^{x \kappa}_{i} \delta \mu_{i} )}$, $\gamma_\kappa={( \sum_{i} T^{y \kappa}_{i} \delta b_{i} )}/\sqrt{e}$, $\delta_\kappa={( \sum_{i} T^{y \kappa}_{i} \delta \nu_{i} )}/\sqrt{e}$.  

 {Moreover, as the system of equations \eqref{wj} and \eqref{hj} is linear, in case of stability, the master stability functions \eqref{eq:MSFn} can be approximated as follows:} 
 \begin{subequations}\label{abcdn}
 \begin{align}
  \mathcal{M}_\kappa= & c_{\alpha 1}(s) |\alpha_\kappa|  +c_{\beta 1}(s) |\beta_\kappa| +c_{\gamma 1}(s) |\gamma_\kappa| + c_{\delta 1}(s) |\delta_\kappa|,\\
       \mathcal{M}_{0}= & c_{\alpha 2} |\alpha_\kappa|  +c_{\beta 2} |\beta_\kappa|,
 \end{align}
\end{subequations}
 {where $c_{\alpha 1},c_{\beta 1},c_{\gamma 1},c_{\delta 1},c_{\alpha 2}, and c_{\beta 2}$ are appropriate coefficients (see also \cite{SOPO}).} 
{Note that Eqs. \eqref{abcdn} relate the master stability functions with the parameter mismatches through the coefficients $\alpha_\kappa,\beta_\kappa,\gamma_\kappa$ and $\delta_\kappa$ defined above.}

{Next, following Refs. \cite{Su:Bo:Ni,SOPO}, we will show how the low-dimensional approach [and the master stability functions defined in \eqref{eq:MSFn}] can be used to quantify the synchronization error observed in the high-dimensional system. The main motivation for performing this analysis is that typically dealing with the low-dimensional systems \eqref{wj} and \eqref{hj} is computationally more convenient than with the high-dimensional system \eqref{eq:SystemOfEqs} (see also Refs. \cite{Su:Bo:Ni,SOPO}). }

We define the synchronization error,
\begin{equation}\label{Elr}
    E(t)=\sqrt{\sum_{i=1}^{N_x} \delta \bx_i^T(t) \delta \bx_i(t)+\sum_{i=1}^{N_y} \delta \by_i^T(t) \delta \by_i(t)}= \sqrt{\sum_{\kappa=2}^{r} {\bp}_\kappa^T(t) {\bp}_\kappa(t)+\sum_{\kappa=1}^{d} \bq_\kappa^T(t) \bq_\kappa(t)}
\end{equation}
 
 Assuming stability, the time average of the synchronization error can be computed in terms of the master stability functions,
 \begin{equation}\label{meanE-MSF}
   <E(t)>_t= \sqrt{\sum_{\kappa=2}^r {\mathcal{M}_\kappa}^2 + \sum_{\kappa=1}^d {\mathcal{M}_0}^2, }
 \end{equation}
 where with the symbol $<...>_t$, we indicate a time-average. Not that $\mathcal{M}_0(0, \alpha_{\kappa}, \beta_{\kappa})$ still depends on $\kappa$ through both arguments $\alpha_{\kappa}$ and $\beta_{\kappa}$, see Eq. \eqref{eq:MSFnb}

\subsection{Spectrum of the matrix $\tilde{M}$}

We have already stated that the singular values of the matrix $\tilde{A}$ can be computed from the eigenvalues of the matrix $\tilde{M}$. We want to show that there is a direct relation between the eigenvalues of the matrix $\tilde{M}$ and those of the
{nominal matrix 
\begin{equation}
M^{NOM} = \left (
         \begin{array}{cc}
           0 & A^{NOM}  \\
           {A^{NOM}}^T & 0
         \end{array}
  \right ),
  \end{equation}
  where here we retain the assumption (with no loss of generality) that the sum of the entries in the rows of the matrix $A^{NOM}$ is equal to $1$. This also implies that the sum of the entries in the columns of the matrix $A^{NOM}$ is equal to $1/e$.
We know from Ref. \cite{NSG} that  the spectrum of the matrix
$M^{NOM}$ is characterized by the following properties: (i) the spectrum is symmetric with respect to the real and imaginary axes, (ii) at least
$d= (N_{x}-N_{y})$ eigenvalues are equal to zero, (iii) two eigenvalues are equal to $+\sqrt{e^{-1}}$ and $-\sqrt{e^{-1}}$ and (iv) the eigenvectors of the matrix $M^{NOM}$,
$\bv^{\kappa}=[ \bv^{x \kappa} \quad  \bv^{y \kappa}]^{T}$ can be of either one of two different types:
\begin{itemize}
\item Type I eigenvectors associated with eigenvalues $\pm \sqrt{e^{-1}}$, $\bv^{1}=[\underbrace{1,1,...,1}_\text{$N_x$ times}, \underbrace{0,0,...,0}_\text{$N_y$ times}]$ and $\bv^{2}=[\underbrace{0,0,...,0}_\text{$N_x$ times}, \underbrace{1,1,...,1}_\text{$N_y$ times}]$. These correspond to modes parallel to the synchronization
manifold.
\item Type II eigenvectors associated with the remaining $(N_{x}+N_{y}-2)$ eigenvalues, for which $\sum_{l=1}^{N_{x}} v^{x \kappa}_l=0$ and $\sum_{l=1}^{N_{y}} v^{y \kappa}_l=0$, $\kappa=3,...,N_x+N_y$. These correspond to modes orthogonal to the synchronization
manifold.
    \end{itemize}

In the first part of this Section we have seen that, in case of parametric mismatches, the master stability function \eqref{eq:MSFn} and the extended master stability function \eqref{abcdn}
depend on the eigenvalues of the matrix $\tilde{M}$ rather than those of the nominal matrix $M^{NOM}$. We want to show that the spectrum of the matrix $\tilde{M}$ is closely related to that of the matrix $M^{NOM}$, namely,
\begin{equation}
    M^{NOM} \pmb{v}^{\kappa} = \varsigma^{\kappa} \pmb{v}^{\kappa} \Leftrightarrow \tilde{M} \pmb{v}^{\kappa} = \vartheta^k \pmb{v}^{\kappa},
\end{equation}
where (a) $\vartheta^1=\vartheta^2=0$ and (b) $\vartheta^\kappa=\varsigma^\kappa$, $\kappa=3,..,N_x+N_y$.

To prove proberty (a) we just need to observe that from the structure of the matrix $\tilde{M}$ in \eqref{E_i} and the property that the sums over the entries in the rows and in the columns of the matrix $\tilde{A}$ is equal to zero, it follows that $\tilde{M} \bv^{\kappa}=\pmb{0}$ for $\kappa=1,2$.  Hence, $\bv^{1}$ and $\bv^{2}$ are still eigenvectors for the matrix $\tilde{M}$ but with associated eigenvalue $0$.

To prove property (b), we see that the left eigenvalue equation for the matrix $\tilde{M}$ corresponds to the following equations: $\bv^{x\kappa^T} \tilde{A}=\varsigma^\kappa \bv^{y\kappa^T}$ and $\bv^{y\kappa^T} \tilde{A}^T=\varsigma^\kappa \bv^{x\kappa^T}$. We then note that from Eq.\ \eqref{TILDE} we can write that the matrix $\tilde{A}=(A^{NOM}-\Delta)$, where the entries over column $j$ of the matrix $\Delta$ are all the same and equal to $a_{j}$, $j=1,...,N_{y}$. Then for $\kappa=3,...,N_{x}+N_{y}$, $\bv^{x\kappa^T} \tilde{A}= \bv^{x\kappa^T} ({A}^{NOM} - \Delta)= \varsigma^\kappa \bv^{y\kappa^T}$ since $\bv^{x\kappa^T}  \Delta =\pmb{0}$ from the property that the sum over the entries of the vector $\bv^{x\kappa}$ is equal to zero, for $\kappa=3,...,N_x+N_y$.
Hence, for $\kappa=3,...,N_x+N_y$, $\bv^{\kappa}$ is still an eigenvector for the matrix $\tilde{M}$ with associated eigenvalue $\varsigma^\kappa$.
}

\subsection{Example}

\begin{figure}[H]
\centering
  \includegraphics[scale=.4]{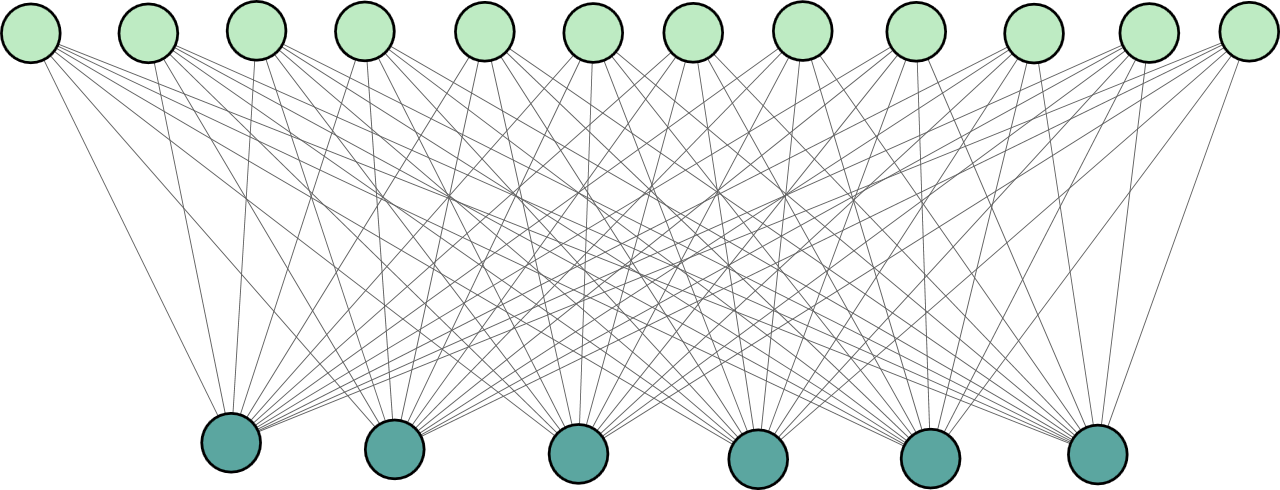}
\caption{A  bipartite network with $N_x=12$ nodes in the $X$ group, shown in light green, and $N_y=6$ nodes in the $Y$ group, shown in dark green. 
The network is bipartite so it contains just inter group connections.}
\label{10n}
\end{figure}

Consider the example of the bipartite network with two groups, $N_x=12, N_y=6$ shown in Fig.\ \ref{10n}.
The matrix $A^{NOM}$ for this network is
\begin{equation} \label{ANOM}
    A^{NOM}=0.2\begin{pmatrix}
0 & 1 & 1 & 1 & 1 & 1 \\ 
0 & 1 & 1 & 1 & 1 & 1 \\ 
1 & 0 & 1 & 1 & 1 & 1 \\ 
1 & 0 & 1 & 1 & 1 & 1 \\ 
1 & 1 & 0 & 1 & 1 & 1 \\ 
1 & 1 & 0 & 1 & 1 & 1 \\ 
1 & 1 & 1 & 0 & 1 & 1 \\ 
1 & 1 & 1 & 0 & 1 & 1 \\ 
1 & 1 & 1 & 1 & 0 & 1 \\ 
1 & 1 & 1 & 1 & 0 & 1 \\ 
1 & 1 & 1 & 1 & 1 & 0 \\ 
1 & 1 & 1 & 1 & 1 & 0 
    \end{pmatrix},
\end{equation}
$r=d=6$.
By calculating the transformation matrix $T$ and following  Eq.\ \eqref{w}, the matrix $\tilde{Q}$ for this network is
\begin{equation}
\tilde{Q}=\frac{0.2}{\sqrt{1/2}} \left(
\begin{array}{cccccccccccccccccc}
\cline{1-2}
\multicolumn{1}{|c}{0} & \multicolumn{1}{c|}{0} & 0 & 0 & 0 & 0 & 0 & 0 & 0 & 0 & 0 & 0 & 0 & 0 & 0 & 0 & 0 & 0 \\ 
\multicolumn{1}{|c}{0} & \multicolumn{1}{c|}{0} & 0 & 0 & 0 & 0 & 0 & 0 & 0 & 0 & 0 & 0 & 0 & 0 & 0 & 0 & 0 & 0 \\ 
\cline{1-4}
0 & 0 & \multicolumn{1}{|c}{0} & \multicolumn{1}{c|}{1} & 0 & 0 & 0 & 0 & 0 & 0 & 0 & 0 & 0 & 0 & 0 & 0 & 0 & 0 \\
0 & 0 & \multicolumn{1}{|c}{1} & \multicolumn{1}{c|}{0} & 0 & 0 & 0 & 0 & 0 & 0 & 0 & 0 & 0 & 0 & 0 & 0 & 0 & 0 \\
\cline{3-6}
0 & 0 & 0 & 0 & \multicolumn{1}{|c}{0} & \multicolumn{1}{c|}{1} & 0 & 0 & 0 & 0 & 0 & 0 & 0 & 0 & 0 & 0 & 0 & 0 \\ 
0 & 0 & 0 & 0 & \multicolumn{1}{|c}{1} & \multicolumn{1}{c|}{0} & 0 & 0 & 0 & 0 & 0 & 0 & 0 & 0 & 0 & 0 & 0 & 0 \\ 
\cline{5-8}
0 & 0 & 0 & 0 & 0 & 0 & \multicolumn{1}{|c}{0} & \multicolumn{1}{c|}{1} & 0 & 0 & 0 & 0 & 0 & 0 & 0 & 0 & 0 & 0 \\ 
0 & 0 & 0 & 0 & 0 & 0 & \multicolumn{1}{|c}{1} & \multicolumn{1}{c|}{0} & 0 & 0 & 0 & 0 & 0 & 0 & 0 & 0 & 0 & 0 \\ 
\cline{7-10}
0 & 0 & 0 & 0 & 0 & 0 & 0 & 0 & \multicolumn{1}{|c}{0} & \multicolumn{1}{c|}{1} & 0 & 0 & 0 & 0 & 0 & 0 & 0 & 0 \\ 
0 & 0 & 0 & 0 & 0 & 0 & 0 & 0 & \multicolumn{1}{|c}{1} & \multicolumn{1}{c|}{0} & 0 & 0 & 0 & 0 & 0 & 0 & 0 & 0 \\ 
\cline{9-12}
0 & 0 & 0 & 0 & 0 & 0 & 0 & 0 & 0 & 0 & \multicolumn{1}{|c}{0} & \multicolumn{1}{c|}{1} & 0 & 0 & 0 & 0 & 0 & 0 \\ 
0 & 0 & 0 & 0 & 0 & 0 & 0 & 0 & 0 & 0 & \multicolumn{1}{|c}{1} & \multicolumn{1}{c|}{0} & 0 & 0 & 0 & 0 & 0 & 0 \\ 
\cline{11-12}
0 & 0 & 0 & 0 & 0 & 0 & 0 & 0 & 0 & 0 & 0 & 0 & 0 & 0 & 0 & 0 & 0 & 0 \\ 
0 & 0 & 0 & 0 & 0 & 0 & 0 & 0 & 0 & 0 & 0 & 0 & 0 & 0 & 0 & 0 & 0 & 0 \\ 
0 & 0 & 0 & 0 & 0 & 0 & 0 & 0 & 0 & 0 & 0 & 0 & 0 & 0 & 0 & 0 & 0 & 0 \\ 
0 & 0 & 0 & 0 & 0 & 0 & 0 & 0 & 0 & 0 & 0 & 0 & 0 & 0 & 0 & 0 & 0 & 0 \\ 
0 & 0 & 0 & 0 & 0 & 0 & 0 & 0 & 0 & 0 & 0 & 0 & 0 & 0 & 0 & 0 & 0 & 0 \\ 
0 & 0 & 0 & 0 & 0 & 0 & 0 & 0 & 0 & 0 & 0 & 0 & 0 & 0 & 0 & 0 & 0 & 0 
\end{array}
\right)
\end{equation}
where each one of the $r$ two-dimensional $Q_i$ blocks is shown in a box. The first box ($Q_1$) corresponds to the parallel motion, with associated singular value $s_1=0$. The other blocks correspond to the transverse motions with associated singular values $s_2=s_3=...=s_6={0.2}/{\sqrt{1/2}}$. By applying the same permutation applied to the matrix $Q$ to the first term on the right hand side of Eq. \eqref{w}, we have:
\begin{equation}
\left(
   \begin{array}{cccccccccccccccccc}
    \cline{1-2}
\multicolumn{1}{|c}{\mathrm{D} \bF_x^*} & \multicolumn{1}{c|}{{0}} & 0 & 0 & 0 & 0 & 0 & 0 & 0 & 0 & 0 & 0 & 0& 0& 0& 0& 0& 0\\ 
\multicolumn{1}{|c}{{0}} & \multicolumn{1}{c|}{\mathrm{D} \bG_y^*} & 0 & 0 & 0 & 0 & 0 & 0 & 0 & 0 & 0 & 0& 0& 0& 0& 0& 0 & 0\\ 
\cline{1-4}
0 & 0 & \multicolumn{1}{|c}{\mathrm{D} \bF_x^*} & \multicolumn{1}{c|}{{0}} & 0 & 0 & 0 & 0 & 0 & 0 & 0 & 0& 0& 0& 0& 0& 0 & 0\\ 
0 & 0 & \multicolumn{1}{|c}{{0}} & \multicolumn{1}{c|}{\mathrm{D} \bG_y^*} & 0 & 0 & 0 & 0 & 0 & 0 & 0 & 0& 0& 0& 0& 0& 0 & 0\\  \cline{3-6}
0 & 0 & 0 & 0 & \multicolumn{1}{|c}{\mathrm{D} \bF_x^*} & \multicolumn{1}{c|}{{0}} & 0 & 0 & 0 & 0 & 0 & 0& 0& 0& 0& 0& 0 & 0\\ 
0 & 0 & 0 & 0 & \multicolumn{1}{|c}{{0}} & \multicolumn{1}{c|}{\mathrm{D} \bG_y^*} & 0 & 0 & 0 & 0 & 0 & 0& 0& 0& 0& 0& 0 & 0\\  \cline{5-8}
0 & 0 & 0 & 0 & 0 & 0 & \multicolumn{1}{|c}{\mathrm{D} \bF_x^*} & \multicolumn{1}{c|}{{0}} & 0 & 0& 0 & 0& 0& 0& 0& 0& 0 & 0\\ 
0 & 0 & 0 & 0 & 0 & 0 & \multicolumn{1}{|c}{{0}} & \multicolumn{1}{c|}{\mathrm{D} \bG_y^*} & 0 & 0& 0 & 0& 0& 0& 0& 0& 0 & 0\\  \cline{7-10}
0 & 0 & 0 & 0 & 0 & 0 & 0 & 0 & \multicolumn{1}{|c}{\mathrm{D} \bF_x^*} & \multicolumn{1}{c|}{{0}}& 0 & 0& 0& 0& 0& 0& 0 & 0\\ 
0 & 0 & 0 & 0 & 0 & 0 & 0 & 0 & \multicolumn{1}{|c}{{0}} & \multicolumn{1}{c|}{\mathrm{D} \bG_y^*} & 0 & 0& 0& 0& 0& 0& 0 & 0\\ 
\cline{9-12}
0 & 0 & 0 & 0 & 0 & 0 & 0 & 0 & 0&0& \multicolumn{1}{|c}{\mathrm{D} \bF_x^*} & \multicolumn{1}{c|}{0} & 0 & 0& 0& 0& 0& 0\\ 
0 & 0 & 0 & 0 & 0 & 0 & 0 & 0 & 0&0& \multicolumn{1}{|c}{{0}} & \multicolumn{1}{c|}{\mathrm{D} \bG_y^*} & 0 & 0& 0& 0& 0& 0\\ 
\cline{11-12}
0 & 0 & 0 & 0 & 0 & 0 & 0 & 0 & 0 & 0 & 0 & 0 & \mathrm{D} \bF_x^*& 0& 0& 0& 0& 0\\ 
0 & 0 & 0 & 0 & 0 & 0 & 0 & 0 & 0 & 0 & 0 & 0 & 0& \mathrm{D} \bF_x^*& 0& 0& 0& 0\\ 
0 & 0 & 0 & 0 & 0 & 0 & 0 & 0 & 0 & 0 & 0 & 0 & 0& 0& \mathrm{D} \bF_x^*& 0& 0& 0\\ 
0 & 0 & 0 & 0 & 0 & 0 & 0 & 0 & 0 & 0 & 0 & 0 & 0& 0& 0& \mathrm{D} \bF_x^*& 0& 0\\ 
0 & 0 & 0 & 0 & 0 & 0 & 0 & 0 & 0 & 0 & 0 & 0 & 0& 0& 0& 0& \mathrm{D} \bF_x^*& 0\\ 
0 & 0 & 0 & 0 & 0 & 0 & 0 & 0 & 0 & 0 & 0 & 0 & 0& 0& 0& 0& 0& \mathrm{D} \bF_x^*\\ 
\end{array}
\right)
\end{equation}

We further note that in case there are no mismatches, that is, $\delta a_i=0$, $\delta b_j=0$, $\delta \mu_i=0$, and $\delta \nu_j=0$, the low-dimensional system \eqref{wj} reduces to the unforced system
\begin{subequations}
\label{eq:lastEq2}
 \begin{align}
     \dot{\tilde{\bx}}(t) &=  \mathrm{D} \bF_x^*\tilde{\bx}(t) + \lambda_{\kappa}  \mathrm{D} \bF_u^* \mathrm{D} \bH_y^* \tilde{\by}(t),  \\
     \dot{\tilde{\by}}(t) &= \mathrm{D} \bG_y^*\tilde{\by}(t) + \lambda_{\kappa}  \mathrm{D} \bG_u^* \mathrm{D} \bL_x^* \tilde{\bx}(t),
 \end{align}
\end{subequations}
 which is the low dimensional solution found in Ref. \cite{NSG}. {Note that Eqs. \eqref{w} is a system of linear time-varying nonhomogeneous differential equations and is characterized by the same stability range as the associated homogeneous system \eqref{eq:lastEq2} \cite{liao2007stability}}. 
 The condition for stability is that the maximum Lyapunov exponents of \eqref{eq:lastEq2} are negative for $\kappa=2,...,r$, excluding $\lambda_1=0$, which is associated with perturbations tangent to the synchronization manifold \cite{NSG}.
 
{ It is important to emphasize that
 Eqs. \eqref{eq:lastEq2} depend on $\bar{x}(t)$ and $\bar{y}(t)$, which are averaged trajectories over all the systems in the network. In a large network, calculating $\bar{x}(t)$ and $\bar{y}(t)$ may be computationally expensive, as it requires full integration of $(N_x+N_y)$ individual
systems, see Eq.\ \eqref{eq:dynamicsCondition}. However, for practical purposes,
$\bar{x}(t)$ and $\bar{y}(t)$ in
\eqref{eq:lastEq2} can be replaced by $x_s(t)$ and $y_s(t)$ obeying Eq.\ \eqref{equs} (see also Ref. \cite{SOPO}).}

    We conclude that in the presence of small parametric mismatches, we can still use \eqref{eq:lastEq2} to determine whether the synchronous solution is stable.
  That also explains the excellent agreement found between the stability region predicted by the theory and experimentally observed in Ref.\ \cite{williams2013experimental}. {While this is not a surprising result, as it is expected that linear stability of a smooth dynamical system is not affected by small parametric variations, by using this approach we will be able to quantify the synchronization error as a function of the parameter mismatches, based on the low-dimensional reduction \eqref{wj} and \eqref{hj}.}

We now consider the following set of equations,

\begin{subequations}\label{L1}
\begin{align}
\dot{x}_{i1}= & -\sigma_i {x_{i1}} +\sigma_i \sum_{j=1}^{N_y} A_{ij} {y_{j}}, \\
\dot{x}_{i2}= & x_{i1} \sum_{j=1}^{N_y} A_{ij} {y_{j}} -\xi x_{i2}
\end{align}
\end{subequations}
$i=1,...,N_x$.

\begin{equation}\label{L2}
\dot{y}_j=-y_j +\sum_{i=1}^{N_x} B_{ji} {x}_{i1}  \Bigl(\rho- \sum_{i=1}^{N_x} B_{ji} x_{i2} \Bigr),
\end{equation}
$j=1,...,N_y$.

We set the nominal parameters as follows:  $\sigma_i=\sigma=10$, $i=1,...,N_x$, $\rho=28$, and  $\xi=2$.
If $\sum_j A_{ij}=1$ and $\sum_i B_{ji}=1$, then the synchronous dynamics corresponds to that of the chaotic Lorenz system \cite{lorenz1963deterministic}.
{Given the set of equations \eqref{L1} and \eqref{L2}, Eqs.\ \eqref{eq:lastEq2} become}

\begin{align*}
\dot{\tilde{x}}_1= & -\sigma \tilde{x}_{1} + \sigma \lambda_{\kappa} \tilde{y}, \\
\dot{\tilde{x}}_2= & \bar{y} \tilde{x}_1 +\bar{x}_1 \lambda_{\kappa} \tilde{y} - \xi {\tilde{x}}_2, \\
\dot{\tilde{y}}  = & -\tilde{y} +  (\rho - \bar{x}_2) \lambda_\kappa \tilde{x}_1 -\bar{x}_1 \lambda_{\kappa} \tilde{x}_2.
\end{align*}

The maximum Lyapunov exponent of this system is plotted versus $\lambda_\kappa$ in Fig.\ \ref{FS}, from which we see that the synchronous solution is stable for $\lambda$ approximately less than $0.435$.

\begin{figure}[t!]
\centering
\includegraphics[scale=.80]{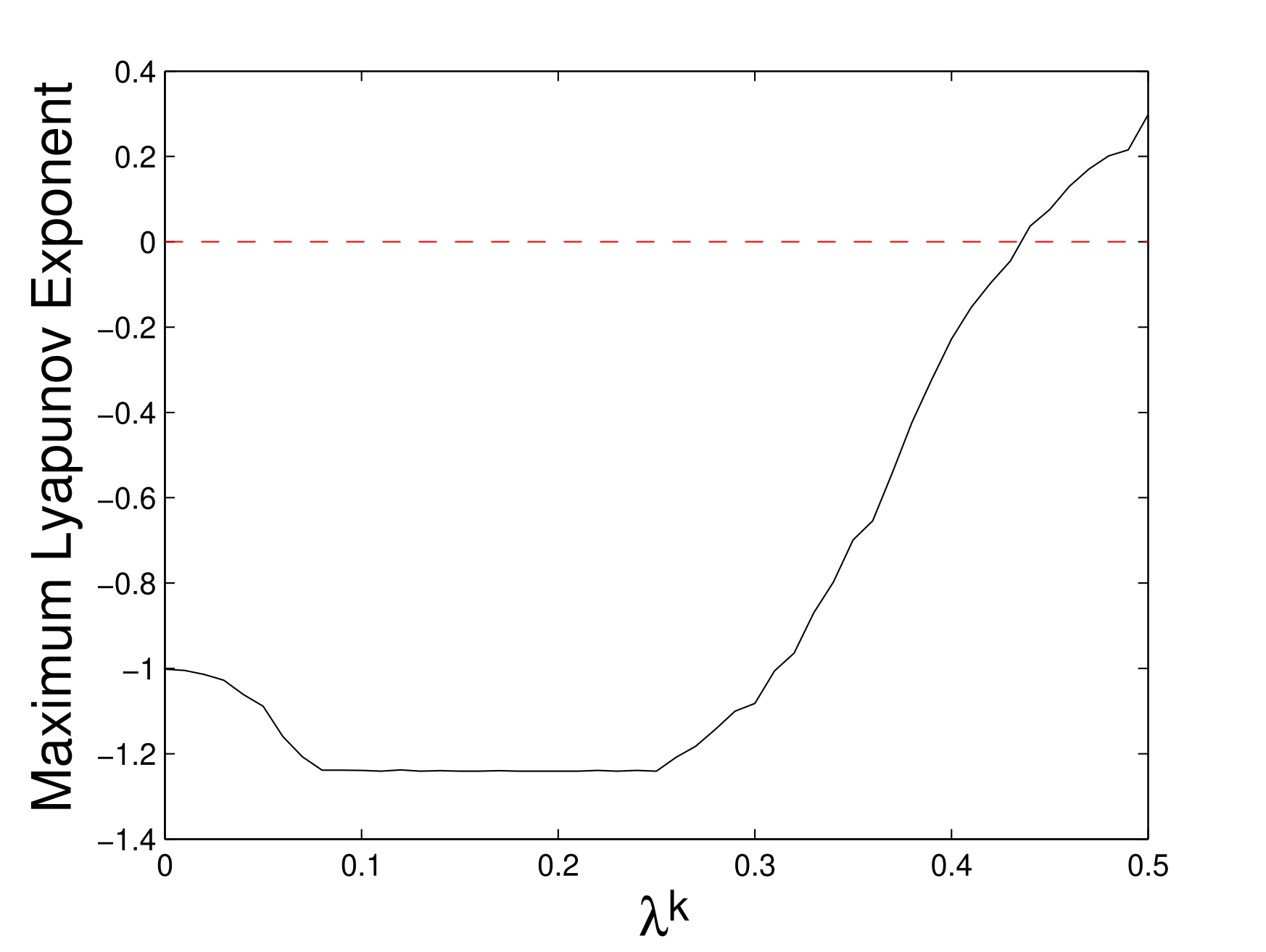}
\caption{Maximum Lyapunov Exponent for the system \eqref{eq:lastEq2} as a function of the parameter $\lambda_k$. The dashed line is the $0$-ordinate line.
}
\label{FS}
\end{figure}

\begin{figure}[H]
\centering
  \includegraphics[scale=.3]{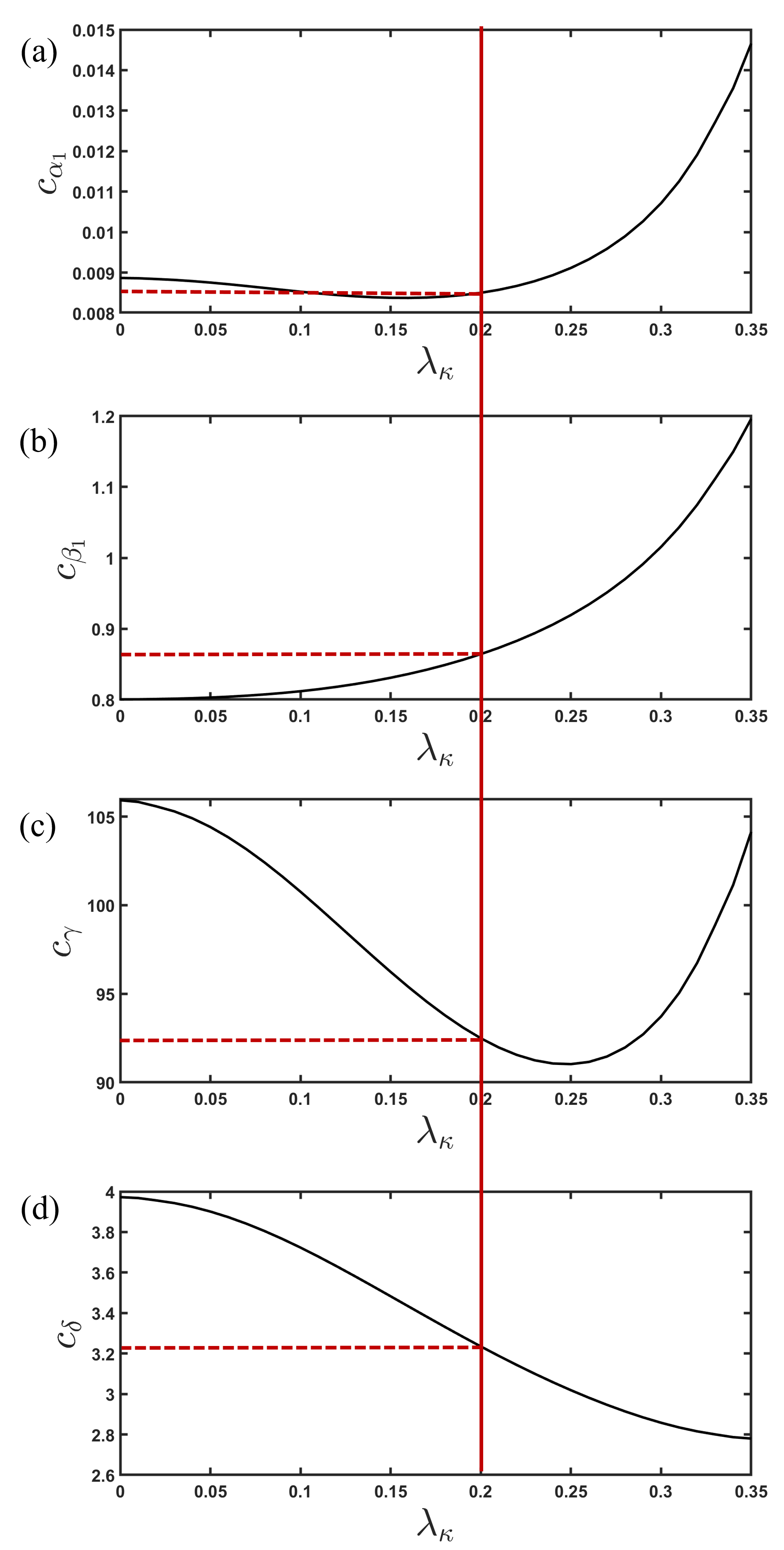}
\caption{$c_{\alpha_{1}}$, $c_{\beta_{1}}$, $c_{\gamma}$, and $c_{\delta}$ versus $\lambda_{\kappa}$. The vertical lines highlight the values of the $c_{\alpha_{1}}$, $c_{\beta_{1}}$, $c_{\gamma}$ and $c_\delta$ for $\lambda_{\kappa}={0.2}.$}
\label{C}
\end{figure}

Figure \ref{C} shows $c_{\alpha_{1}}$, $c_{\beta_{1}}$, $c_{\gamma}$, and $c_{\delta}$ as functions of $\lambda_{\kappa}$. The two constant values $c_{\alpha_{2}}=51.7524$ and $c_{\beta_{2}}=0.8002$.
Using this information and Eqs.\ \eqref{abcdn}, we can compute the master
stability function as a function of $\lambda_{\kappa}$. This in turn allows us to approximate the synchronization error, using Eq.\ \eqref{meanE-MSF} from knowledge of $\alpha_\kappa$, $\beta_\kappa$, $\gamma_\kappa$, and $\delta_\kappa$, as long as all the $\lambda_\kappa$'s are in the region of stability, $0 \leq \lambda_\kappa \leq 0.435$. We wish to emphasize that this result is general and applies to any bipartite network with bidirectional connections, whose time-evolution obeys Eqs.\ \eqref{L1} and \eqref{L2}.

In order to validate the theory, we consider the set of Eqs.\ \eqref{L1} and \eqref{L2} for the network topology in Fig.\ \ref{10n} [$A^{NOM}$ in Eq.\ \eqref{ANOM}] and study the following two cases:
\begin{enumerate}
    \item[(I)] Small mismatches affecting the couplings, namely $A_{ij}=A_{ij}^{NOM}(1+5 \times 10^{-5} \phi_{ij})$, where $\phi_{ij}$ is a random number from a standard normal distribution. \\
    \item[(II)] Small mismatches affecting both the couplings and  the individual system parameters,  namely  $A_{ij}=A_{ij}^{NOM}(1+ 5 \times 10^{-5} \phi_{ij})$ and $\sigma_i$ in Eq.\ (\ref{L1}a) is given by $\sigma_i=10(1+ 10^{-6} \psi_i)$, where both $\phi_{ij}$ and $\psi_i$ are random numbers from a standard normal distribution.  
\end{enumerate}
{All the transverse singular values of the matrix $A^{NOM}$ in Eq.\ \eqref{ANOM} are equal to $s_\kappa={0.2}/{\sqrt{1/2}}$. To these corresponds only one value of $\lambda_\kappa=0.2$.}  Table \ref{TableE} compares the synchronization error obtained from integration of the full high-dimensional system \eqref{L1} and \eqref{L2} [left hand side of Eq.\ \eqref{eq:MSFn}] and the estimated synchronization error using the master stability function [right hand side of Eq.\ \eqref{eq:MSFn}], for both cases described above.

\begin{table}[h]
    \centering
\caption{The synchronization error observed in the high-dimensional system and the estimated synchronization error using the master stability function defined in Eq. \eqref{eq:MSFn} for two different cases. }
\setlength{\tabcolsep}{10pt} 
\renewcommand{\arraystretch}{1.5} 
\begin{tabular}{ |c|c|c|} 
\hline
\shortstack{Case} & \shortstack{$<E(t)>_t$} & \shortstack{$\sqrt{\sum_{\kappa=2}^r {\mathcal{M}_\kappa}^2 + \sum_{\kappa=1}^d {\mathcal{M}_0}^2}$}\\
\hline
(I) & $1.529\times10^{-3}$ & $1.448\times10^{-3}$ \\
(II) & $1.609 \times10^{-3}$ & $1.614\times10^{-3}$ \\
\hline
\end{tabular}
\label{TableE}
\end{table}

\color{black}

\section{Couplings among members of the same group}\label{III}

So far we have only considered coupling from the $\bX$ group to the $\bY$ group and vice versa. While in general group synchronization is possible in the presence of intra group connections, stability of the group synchronous solution when also intra group connections are present has not been fully elucidated. This problem was investigated in Ref. \cite{clustergroup}, where conditions were presented for the dimensionality reduction of the stability problem. Here we will extend some of the results in Ref. \cite{clustergroup} and we will show that the stability of group synchronization can be reduced in a low-dimensional form for a broader class of networks.

Consider two groups of coupled oscillators, described by the following equations:
\begin{subequations}\label{IG}
 \begin{align}
      \dot{\bx}_{i} &= F(\bx_{i} ) + \sum_{j=1}^{N_{y}}A_{ij}H(\by_{j}) +  \sum_{\ell=1}^{N_{x}} C_{i\ell} R(\bx_{\ell}), \quad  i=1,\hdots,N_{x}, \\
      \dot{\by}_{j} &= G(\by_{j} ) + \sum_{i=1}^{N_{x}}B_{ji}L(\by_{i}) +  \sum_{\ell=1}^{N_{y}} D_{j\ell} S(\by_{\ell}), \quad  j=1,\hdots,N_{y},
 \end{align}
\end{subequations}
where the output functions $R: \mathbf{R}^{n_x} \rightarrow  \mathbf{R}^{n_x}$ and $S: \mathbf{R}^{n_y} \rightarrow  \mathbf{R}^{n_y}$ and the $N_x$-square matrix $C$ and the $N_y$-square matrix $D$ define the intra group connections of group $\bX$ and group $\bY$, respectively.

If $\sum_{j=1}^{N_{y}} A_{ij}= a \neq 0$ and $\sum_{\ell=1}^{N_x} C_{i\ell} =c \neq 0$ independent of $i$ and $\sum_{i}^{N_{x}} B_{ji}= b \neq 0$ and $\sum_{\ell=1}^{N_{y}} D_{j\ell}=d \neq 0$ independent of $j$, a synchronous solution exists $x_{1}=x_{2}=...=x_{N_{x}}=x_{s}$,
$y_{1}=y_{2}=...=y_{N_{y}}=y_{s}$, obeying,
\begin{subequations}
	\begin{align}
  \dot{\bx}_{s} & = F(\bx_{s}) + a H(\by_{s})+ c R(\bx_s)  \\
  \dot{\by}_{s} & = G(\by_{s}) + b L(\bx_{s})+ d S(\by_s).
  \end{align}
\end{subequations}
Note that by appropriately rescaling the functions $\pmb{H},\pmb{L},\pmb{R},\pmb{S}$ it is always possible to set $a=b=c=d=1$. Therefore, without loss of generality, in what follows we will proceed under this assumption.

In order to analyze stability, we linearize \eqref{IG} about the synchronous solution,
\begin{subequations} \label{linIG}
	\begin{align}
		\delta \dot{\bx}_{i} &= DF(\bx_{s})\delta \bx_{i} + \sum_{j=1}^{N_{y}} A_{ij}DH(\by_{s}) \delta \by_{j} + \sum_{\ell=1}^{N_{x}} C_{i\ell} DR(\bx_{s}) \delta \bx_{\ell}, \quad i=1,...,N_x\\
		\delta \dot{\by}_{j} &= DG(\by_{s})\delta \by_{j} + \sum_{i=1}^{N_{y}} B_{ji}DL(\bx_{s}) \delta \bx_{i} + \sum_{\ell=1}^{N_{y}} D_{j\ell} DS(\by_{s}) \delta \by_{\ell}, \quad j=1,...,N_y.
	\end{align}
\end{subequations}

In \cite{clustergroup} it was shown that Eq.\ \eqref{linIG} can be reduced in a low-dimensional form, provided that the matrix $N=\begin{bmatrix}
		C	&	0 	\\
		0 	& 	D
	\end{bmatrix}$ has the same set of eigenvectors as the matrix  $M=\begin{bmatrix}
		0	&	A 	\\
		B 	& 	0
	\end{bmatrix}$. It was thus concluded in \cite{clustergroup} that if this condition is satisfied, in a similar way to the case of bipartite topologies, stability of the synchronous solution can be reduced in the following low dimensional form,
\begin{subequations}\label{vld}
	\begin{align}
\dot{\bar{\bx}}^{\kappa} &= DF(\bx_{s})  \bar{\bx}^{\kappa} +  {\lambda^{\kappa}} DH(\by_{s})  \bar{\by}^{\kappa} + \eta^{\kappa} DR(\bx_s)\bar{\bx}^{\kappa}   \\
		 \dot{\bar{\by}}^{\kappa} &= DG(\by_{s})  \bar{\by}^{\kappa}+ \lambda^{\kappa} DL(\bx_{s})  \bar{\bx}^{\kappa} + \eta^{\kappa} DS(\by_s) \bar{\by}^{\kappa},
	\end{align}
\end{subequations}
$\kappa=1,...,N_x+N_y$, where $\lambda^{\kappa}$ and $\eta^{\kappa}$ are the eigenvalues associated with the same eigenvector of the matrices $M$ and $N$, respectively. By construction, there are always two $\kappa$ eigenvectors which are associated with perturbations tangent to the synchronization manifold, and the corresponding pairs of eigenvalues $(\lambda^{\kappa},\eta^{\kappa})$ do not need to be considered in order to assess stability. Without loss of generality, we label  $\kappa=1$ and $\kappa=2$ the two {tangent eigenmodes} and $\kappa=3,...,N_x+N_y$ the remaining {transverse eigenmodes}.

In what follows we extend the results in \cite{clustergroup} and show that a reduction in a low dimensional form is possible for any pair of matrices
\begin{subequations} \label{CD}
\begin{align}
C=C'+J^x, \\
D=D'+J^y.
\end{align}
\end{subequations}
where (i) the matrix $\begin{bmatrix}
		C'	&	0 	\\
		0 	& 	D'
	\end{bmatrix}$   has the same set of eigenvectors as the matrix  $M$ and (ii) $J^x$ and $J^y$ are any two $N_x$-square and $N_y$-square matrices whose columns are composed of entries that are all the same. {This form of the matrices $J^x$ and $J^y$ corresponds to a particular coupling configuration for which for any pair of nodes $(i,j)$ in either group $\bX$ or group $\bY$, the coupling strength from node $i$ to node $j$ is a function of $i$ but not of $j$.}

Using the ansatz $\delta \bx_{i}(t)=p_{i}^\kappa \hat{\bx}^\kappa(t)$ and $\delta \by_{j}(t)=q_{j}^\kappa \hat{\by}^\kappa(t)$ in \eqref{linIG}, we obtain,
\begin{subequations}\label{eq:system_problemTerm}
	\begin{align}
		{p_{i}^{\kappa}} \dot{\hat{\bx}}^{\kappa} &= DF(\bx_{s}) {p_{i}^{\kappa}} \hat{\bx}^{\kappa} + \sum_{j=1}^{N_{y}} A_{ij} {q_{j}^{\kappa}} DH(\by_{s})  \hat{\by}^{\kappa} + {\sum_{\ell=1}^{N_{x}} C_{i\ell} DR(\bx_s) p_{\ell}^{\kappa} }  \hat{\bx}^{\kappa} \\
		{q_{j}^{\kappa}} \dot{\hat{\by}}^{\kappa} &= DG(\by_{s}) {q_{j}^{\kappa}} \hat{\by}^{\kappa}+ \sum_{i=1}^{N_{x}} B_{ji} {p_{i}^{\kappa}} DL(\bx_{s})  \hat{\bx}^{\kappa} + {\sum_{\ell=1}^{N_{y}} D_{j\ell} DS(\by_s) q_{\ell}^{\kappa} } \hat{\by}^{\kappa}
	\end{align}
\end{subequations}

We now consider the left eigenvalue equation,
\begin{equation} \label{left_eig}
\pmb{\omega}^{\kappa^{T}} M= \lambda^{\kappa}\pmb{\omega}^{\kappa^{T}},
\end{equation}
where the eigenvectors
$\pmb{\omega}^{\kappa^{T}} = \Big( p_{1}^{\kappa^{\mathit{l}}},p_{2}^{\kappa^{\mathit{l}}},\dots,p_{N_{x}}^{\kappa^{\mathit{l}}},q_{1}^{\kappa^{\mathit{l}}},q_{2}^{\kappa^{\mathit{l}}},\dots,q_{N_{y}}^{\kappa^{\mathit{l}}} \Big)$.
By multiplying (\ref{eq:system_problemTerm}a) by $p_{i}^{\kappa^{\mathit{l}}}$, summing over $i$, and dividing by $(\sum_{i} p_i^{\kappa^{\mathit{l}}} {p_i^\kappa})$  [by multiplying (\ref{eq:system_problemTerm}b)
by $q_{j}^{\kappa^{\mathit{l}}}$, summing over $j$, and dividing by $( \sum_{j} q_j^{\kappa^{\mathit{l}}} {q_j^\kappa} )$], we obtain,
\begin{subequations}\label{left}
	\begin{align}
\dot{\hat{\bx}}^{\kappa} &= DF(\bx_{s})  \hat{\bx}^{\kappa} +  {\lambda^{\kappa}} DH(\by_{s})  \hat{\by}^{\kappa} +  \eta^{\kappa} DR(\bx_s) \hat{\bx}^{\kappa} + \frac{\Big( \sum_{\ell} \sum_{i} p_{i}^{\kappa^{\mathit{l}}} J^x_{i\ell} p_{\ell}^{\kappa} \Big)}{\Big(\sum_{i} {p_i^{\kappa^{\mathit{l}}}} {p_i^\kappa} \Big)} DR(\bx_s)\hat{\bx}^{\kappa},  \\
		 \dot{\hat{\by}}^{\kappa} &= DG(\by_{s})  \hat{\by}^{\kappa}+ \lambda^{\kappa} DL(\bx_{s})  \hat{\bx}^{\kappa}+  \eta^{\kappa} DS(\by_s) \hat{\by}^{\kappa} + \frac{ \Big( \sum_{\ell} \sum_{j} q_{j}^{\kappa^{\mathit{l}}} J^y_{j\ell} q_{\ell}^{\kappa} \Big)}{\Big(\sum_{j} {q_j^{\kappa^{\mathit{l}}}} {q_j^\kappa} \Big)} DS(\by_s) \hat{\by}^{\kappa} ,
	\end{align}
\end{subequations}
$\kappa=1,...,N_x+N_y$.

{We shall now show that $\sum_{i} p_{i}^{\kappa^{\mathit{l}}} J^x_{i\ell}=0$, for $\kappa=3,...,(N_x+N_y)$ in Eq.\ (\ref{left}a) and $\sum_{j} q_{j}^{\kappa^{\mathit{l}}} J^y_{j\ell}=0$, for $\kappa=3,...,(N_x+N_y)$ in Eq.\ (\ref{left}b). Since by definition the matrices $J^x$ and $J^y$ have columns whose entries are all the same, it suffices to show that $\sum_{i} p_{i}^{\kappa^{\mathit{l}}}=0$, for $\kappa=3,...,(N_x+N_y)$ and $\sum_{j} q_{j}^{\kappa^{\mathit{l}}}=0$, for $\kappa=3,...,(N_x+N_y)$. We will prove that this latter property is indeed satisfied by the $(N_x+N_y-2)$ eigenvectors that are associated with the transverse eigenmodes of the matrix $M$.
}

{Note that the left eigenvalue equation $\pmb{\omega}^{\kappa^{T}} {M} = \lambda^{\kappa}\pmb{\omega}^{\kappa^{T}}$ implies the following two equations: $\pmb{q}^{\kappa^{T}} {B} = \lambda^{\kappa} \pmb{p}^{\kappa^{T}}$ and $\pmb{p}^{\kappa^{T}} {A} = \lambda^{\kappa} \pmb{q}^{\kappa^{T}}$ where the vectors $\pmb{p}^{\kappa^{T}} = \Big( p_{1}^{\kappa^{\mathit{l}}},p_{2}^{\kappa^{\mathit{l}}},\dots,p_{N_{x}}^{\kappa^{\mathit{l}}}\Big)$ and $\pmb{q}^{\kappa^{T}} = \Big(q_{1}^{\kappa^{\mathit{l}}},q_{2}^{\kappa^{\mathit{l}}},\dots,q_{N_{y}}^{\kappa^{\mathit{l}}} \Big)$, or equivalently,
\begin{subequations}\label{extra}
	\begin{align}
\sum_{j=1}^{N_y} q_j^{\kappa^{\mathit{l}}} {B}_{ji} &= \lambda^{\kappa} p_i^{\kappa^{\mathit{l}}} \\
\sum_{i=1}^{N_x} p_i^{\kappa^{\mathit{l}}} {A}_{ij} &= \lambda^{\kappa} q_j^{\kappa^{\mathit{l}}}.
	\end{align}
\end{subequations}
By summing Eq.\ (\ref{extra}a) over $i=1,...,N_x$  and Eq.\ (\ref{extra}b) over $j=1,...,N_y$, we obtain
\begin{subequations}\label{extra2}
	\begin{align}
\sum_{j=1}^{N_y} q_j^{\kappa^{\mathit{l}}}  &= \lambda^{\kappa}  \sum_{i=1}^{N_x} p_i^{\kappa^{\mathit{l}}} \\
\sum_{i=1}^{N_x} p_i^{\kappa^{\mathit{l}}}  & = \lambda^{\kappa} \sum_{j=1}^{N_y} q_j^{\kappa^{\mathit{l}}}.
	\end{align}
\end{subequations}
By plugging (\ref{extra2}b) into (\ref{extra2}a) we obtain $\sum_{j=1}^{N_y} q_j^{\kappa^{\mathit{l}}}  = (\lambda^{\kappa})^2  \sum_{j=1}^{N_y} q_j^{\kappa^{\mathit{l}}}$, which can be satisfied in two possible ways: either $\lambda^{\kappa}= \pm 1$, which applies to the parallel eigenmodes $\kappa=1,2$ of the matrix $M$ or $\sum_{j=1}^{N_y} q_j^{\kappa^{\mathit{l}}}=0$, which applies to the transverse eigenmodes $\kappa=3,...,(N_x+N_y)$ of the matrix $M$. Analogously, by plugging (\ref{extra2}a) into (\ref{extra2}b), we can show that $\sum_{i=1}^{N_x} p_j^{\kappa^{\mathit{l}}}=0$ for $\kappa=3,...,(N_x+N_y)$.
}

 We can thus conclude that Eq.\ \eqref{left} reduces to \eqref{vld} for $\kappa=3,...,(N_x+N_y)$ [while in general it does not reduce to \eqref{vld} for $\kappa=1,2$]. This implies that when intra group connections are in the general form of Eq.\ \eqref{CD}, transverse stability is still determined by the low-dimensional system  \eqref{vld}.

Finally,  we note that the analysis of parameter mismatches presented in Sec.\ref{II} can be extended to the case that small mismatches affect the intra group connections. In particular, if the intra group connections are affected by small mismatches but are in a form that approximately satisfies  Eq.\ \eqref{CD}, then the low-dimensional system  \eqref{vld}  can still be used to predict stability of the synchronous state.

\subsection{Numerical Example}

{As a numerical example, we considered a small network for which $N_x=2$, $N_y=3$, and
}\begin{equation}
A= \left(
     \begin{array}{ccc}
       w & 1-w & 0\\
       0 & 1-w & w \\
     \end{array}
   \right), \quad B= \left(
                       \begin{array}{cc}
                         1 & 0 \\
                         1 \over 2 & 1 \over 2 \\
                         0 & 1 \\
                       \end{array}
                     \right).
\end{equation}
{In Ref. \cite{NSG} we showed that for this choice of the coupling matrices $A$ and $B$, the eigenvalues $\lambda^{\kappa}$ were equal to $0, \pm \sqrt{w}, \pm 1$. Then, by setting ${w}=0.64$ stability is only affected by the pair of eigenvalues $\lambda^{\kappa}$, $0$ and $\sqrt{w}=0.8$ \cite{NSG}. In our numerical experiments, we retained the set of equation \eqref{L1} and replaced \eqref{L2} by the following:
\begin{equation}\label{L3}
\dot{y}_j=-y_j +\sum_{i=1}^{N_x} B_{ji} {x}_{i1}  \Bigl(\rho- \sum_{i=1}^{N_x} B_{ji} x_{i2} \Bigr) +  \sum_{\ell=1}^{N_{y}} D_{j\ell} S(y_{\ell}),
\end{equation}
$j=1,...,N_y$, where the last term on the right-hand-side of \eqref{L3} represents coupling between systems of the same group $\bY$ and we set the coupling function $S(y_{\ell})= s y_{\ell}$, with $s$ being a variable scalar quantity. }

\begin{figure}[t!]
\centering
\includegraphics[scale=.80]{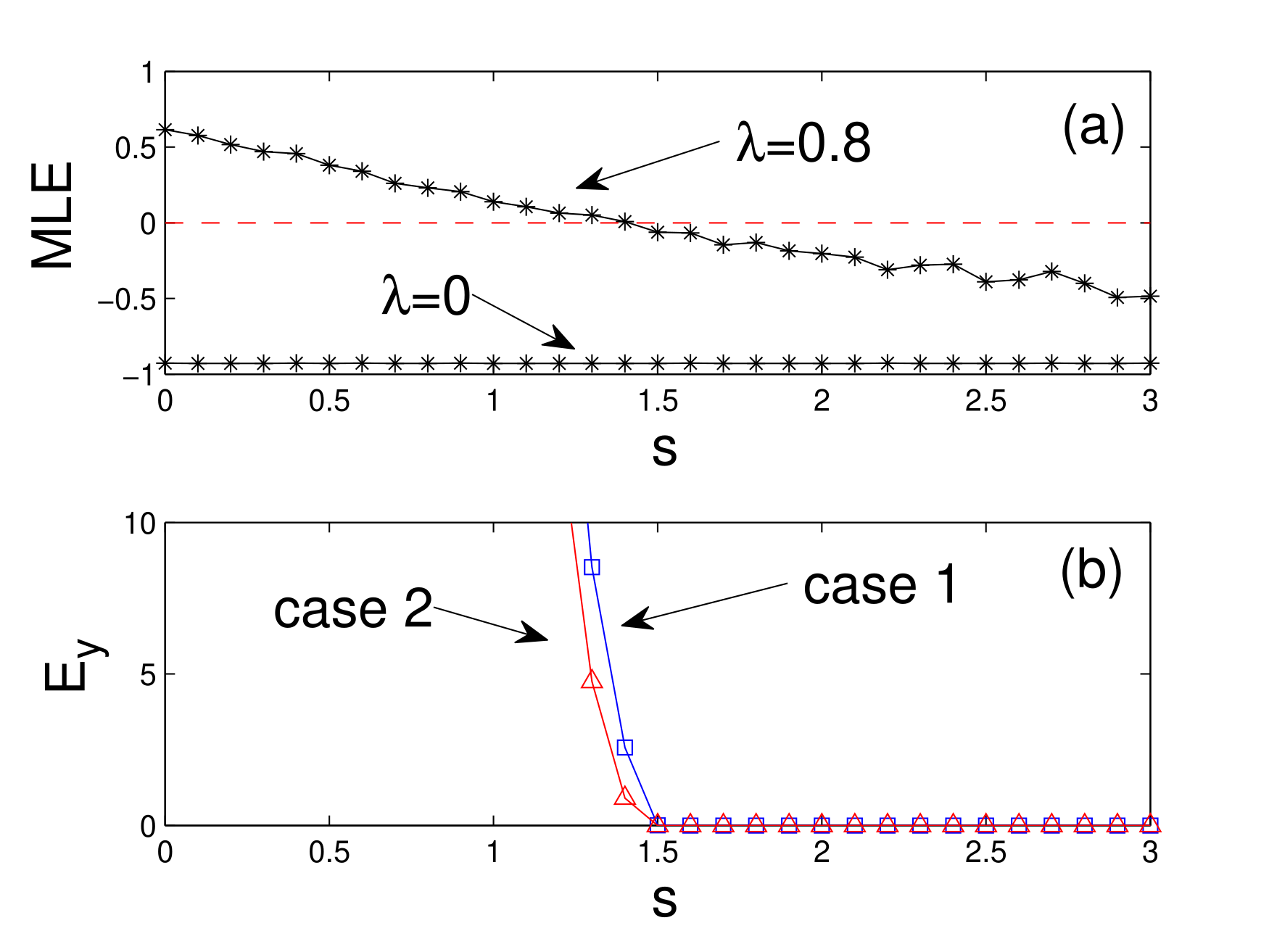}
\caption{ (a) Maximum Lyapunov Exponent (MLE) for the system \eqref{vld} vs. $s$ for $\lambda^{\kappa}=0$ and $\lambda^{\kappa}=0.8$. The dashed line corresponds to the $0$-ordinate line. (b) Synchronization error $E_y$ vs. $s$ for the two choices of the matrix $D$ in Eq.\ \eqref{D}.}
\label{FS}   
\end{figure}

{We considered the following two cases of intragroup coupling matrices $D$,}
\begin{equation} \label{D}
 \left(
     \begin{array}{ccc}
       1/3 & 1/3 & 1/3\\
       1/3 & 1/3 & 1/3 \\
       1/3 & 1/3 & 1/3
     \end{array}
   \right), \qquad \left(
                       \begin{array}{ccc}
       0 & 1 & 0\\
       0 & 1 & 0 \\
       0 & 1 & 0
     \end{array}
                     \right),
\end{equation}
{which we label case 1 and case 2, respectively. For both cases 1 and 2, the corresponding matrices $M$ and $N$ do not commute, so neither case can be studied within the framework presented in Ref. \cite{clustergroup}. Note that both choices of $D$ above are matrices whose columns are composed of entries that are all the same. Therefore, in both cases we expect stability to be described by
the low-dimensional system \eqref{vld} with $\eta^{\kappa}=0$. Fig. \ref{FS}(a) shows the Maximum Lyapunov Exponent (MLE) for the system \eqref{vld} as a function of the parameter $s$ for $\lambda^{\kappa}=0$ and $\lambda^{\kappa}=0.8$, from which we see that stability is expected for $s \gtrsim 1.5$.}

{In Fig.\ref{FS}(b) we run full numerical simulations of the network dynamics [Eqs.\ \eqref{L1} and \eqref{L3}] as we vary the  parameter $s$. For each run, the network systems are evolved from initial conditions that are close to the synchronization manifold. The figure shows the final synchronization error, 
\begin{equation}\label{Ey}
E_y= N_y^{-1} (\tau)^{-1} \int_{t_1}^{t_1+\tau} \sum_{i=1}^{N_y} \Bigl(y_i(t)-\bar{y}(t) \Bigr)^2 dt,
\end{equation}
 with $t_1$ a large-enough time past the initial transient dynamics and $\tau$ a large-enough averaging window,
versus $s$. As can be seen, perfect agreement is attained with the low-dimensional predictions of Fig.\ \ref{FS}(a).}

\section{Conclusions}

The main contribution of this paper is to extend the analysis of the effects of parameter mismatches on the stability of the network synchronous solution \cite{restr_bubbl,Su:Bo:Ni,SOPO,cho2019concurrent, sorrentino2016approximate,acharyya2012synchronization,acharyya2015synchronization}
to the case of networks formed of systems of different types. {An important reference for our work is Ref.  \cite{Su:Bo:Ni} which introduced a master stability function to characterize the synchronization error in networks with parameter mismatches.}

Group synchronization was first studied in \cite{NSG} and has been the subject of both theoretical \cite{clustergroup} and experimental \cite{williams2013experimental} investigations. Group synchronization has been observed and characterized when the systems in each group are identical and the couplings between the systems satisfy specific conditions. In this paper, we have defined a master stability function that describes stability of the group synchronization solution in the presence of mismatches on the individual parameters of the network oscillators and on the couplings between the oscillators in the groups.  This is relevant to experimental realizations of group synchrony  \cite{williams2013experimental}, for which parameter mismatches are practically unavoidable. {Our analysis applies to the case that for each connection from node $i$ in group $\bX$ to node $j$ in group $\bY$, there is a connection from node $j$ in group $\bY$ to node $i$ in group $\bX$ and vice versa.}
We have also extended the analysis presented in Ref. \cite{clustergroup} to study stability of the group synchronization solution in the presence of intra group couplings. Our analysis has pointed out a broader class of matrices describing intra group connectivity for which the stability problem can be reduced in a low-dimensional form.

\section*{Acknowledgement}
The authors thank Prof. Kazuo Murota for insightful discussions on the subject of $*$-algebra. This research is supported by NIH grant 1R21EB028489-01A1.


\begin{thebibliography}{23}
\expandafter\ifx\csname natexlab\endcsname\relax\def\natexlab#1{#1}\fi
\expandafter\ifx\csname bibnamefont\endcsname\relax
  \def\bibnamefont#1{#1}\fi
\expandafter\ifx\csname bibfnamefont\endcsname\relax
  \def\bibfnamefont#1{#1}\fi
\expandafter\ifx\csname citenamefont\endcsname\relax
  \def\citenamefont#1{#1}\fi
\expandafter\ifx\csname url\endcsname\relax
  \def\url#1{\texttt{#1}}\fi
\expandafter\ifx\csname urlprefix\endcsname\relax\def\urlprefix{URL }\fi
\providecommand{\bibinfo}[2]{#2}
\providecommand{\eprint}[2][]{\url{#2}}

\bibitem[{\citenamefont{Arenas et~al.}(2008)\citenamefont{Arenas, Diaz-Guilera,
  Kurths, Moreno, , and Zhou}}]{SReport}
\bibinfo{author}{\bibfnamefont{A.}~\bibnamefont{Arenas}},
  \bibinfo{author}{\bibfnamefont{A.}~\bibnamefont{Diaz-Guilera}},
  \bibinfo{author}{\bibfnamefont{J.}~\bibnamefont{Kurths}},
  \bibinfo{author}{\bibfnamefont{Y.}~\bibnamefont{Moreno}}, , \bibnamefont{and}
  \bibinfo{author}{\bibfnamefont{C.}~\bibnamefont{Zhou}},
  \bibinfo{journal}{Phys. Rep.} \textbf{\bibinfo{volume}{469}},
  \bibinfo{pages}{93} (\bibinfo{year}{2008}).

\bibitem[{\citenamefont{Pecora and Carroll}(1998)}]{Pe:Ca}
\bibinfo{author}{\bibfnamefont{L.}~\bibnamefont{Pecora}} \bibnamefont{and}
  \bibinfo{author}{\bibfnamefont{T.}~\bibnamefont{Carroll}},
  \bibinfo{journal}{Phys. Rev. Lett.} \textbf{\bibinfo{volume}{80}},
  \bibinfo{pages}{2109} (\bibinfo{year}{1998}).

\bibitem[{\citenamefont{Sorrentino and Ott}(2007)}]{NSG}
\bibinfo{author}{\bibfnamefont{F.}~\bibnamefont{Sorrentino}} \bibnamefont{and}
  \bibinfo{author}{\bibfnamefont{E.}~\bibnamefont{Ott}},
  \bibinfo{journal}{Phys. Rev. E} \textbf{\bibinfo{volume}{76}},
  \bibinfo{pages}{056114} (\bibinfo{year}{2007}).

\bibitem[{\citenamefont{Dahms et~al.}(2012)\citenamefont{Dahms, Lehnert, and
  Sch{\"o}ll}}]{clustergroup}
\bibinfo{author}{\bibfnamefont{T.}~\bibnamefont{Dahms}},
  \bibinfo{author}{\bibfnamefont{J.}~\bibnamefont{Lehnert}}, \bibnamefont{and}
  \bibinfo{author}{\bibfnamefont{E.}~\bibnamefont{Sch{\"o}ll}},
  \bibinfo{journal}{Physical Review E} \textbf{\bibinfo{volume}{86}},
  \bibinfo{pages}{016202} (\bibinfo{year}{2012}).

\bibitem[{\citenamefont{Williams et~al.}(2013)\citenamefont{Williams, Murphy,
  Roy, Sorrentino, Dahms, and Sch{\"o}ll}}]{williams2013experimental}
\bibinfo{author}{\bibfnamefont{C.~R.} \bibnamefont{Williams}},
  \bibinfo{author}{\bibfnamefont{T.~E.} \bibnamefont{Murphy}},
  \bibinfo{author}{\bibfnamefont{R.}~\bibnamefont{Roy}},
  \bibinfo{author}{\bibfnamefont{F.}~\bibnamefont{Sorrentino}},
  \bibinfo{author}{\bibfnamefont{T.}~\bibnamefont{Dahms}}, \bibnamefont{and}
  \bibinfo{author}{\bibfnamefont{E.}~\bibnamefont{Sch{\"o}ll}},
  \bibinfo{journal}{Phys. Rev. Lett.} \textbf{\bibinfo{volume}{110}},
  \bibinfo{pages}{064104} (\bibinfo{year}{2013}).

\bibitem[{\citenamefont{Yang}(2015)}]{yang2015group}
\bibinfo{author}{\bibfnamefont{Y.~F.} \bibnamefont{Yang}}, in
  \emph{\bibinfo{booktitle}{Applied Mechanics and Materials}}
  (\bibinfo{organization}{Trans Tech Publ}, \bibinfo{year}{2015}), vol.
  \bibinfo{volume}{733}, pp. \bibinfo{pages}{902--905}.

\bibitem[{\citenamefont{Panahi et~al.}(2021)\citenamefont{Panahi, Klickstein,
  and Sorrentino}}]{panahi2021cluster}
\bibinfo{author}{\bibfnamefont{S.}~\bibnamefont{Panahi}},
  \bibinfo{author}{\bibfnamefont{I.}~\bibnamefont{Klickstein}},
  \bibnamefont{and}
  \bibinfo{author}{\bibfnamefont{F.}~\bibnamefont{Sorrentino}},
  \bibinfo{journal}{arXiv preprint arXiv:2109.13792}  (\bibinfo{year}{2021}).

\bibitem[{\citenamefont{Della~Rossa et~al.}(2020)\citenamefont{Della~Rossa,
  Pecora, Blaha, Shirin, Klickstein, and Sorrentino}}]{della2020symmetries}
\bibinfo{author}{\bibfnamefont{F.}~\bibnamefont{Della~Rossa}},
  \bibinfo{author}{\bibfnamefont{L.}~\bibnamefont{Pecora}},
  \bibinfo{author}{\bibfnamefont{K.}~\bibnamefont{Blaha}},
  \bibinfo{author}{\bibfnamefont{A.}~\bibnamefont{Shirin}},
  \bibinfo{author}{\bibfnamefont{I.}~\bibnamefont{Klickstein}},
  \bibnamefont{and}
  \bibinfo{author}{\bibfnamefont{F.}~\bibnamefont{Sorrentino}},
  \bibinfo{journal}{Nature communications} \textbf{\bibinfo{volume}{11}},
  \bibinfo{pages}{1} (\bibinfo{year}{2020}).

\bibitem[{\citenamefont{Restrepo et~al.}(2004)\citenamefont{Restrepo, Ott, and
  Hunt}}]{restr_bubbl}
\bibinfo{author}{\bibfnamefont{J.~G.} \bibnamefont{Restrepo}},
  \bibinfo{author}{\bibfnamefont{E.}~\bibnamefont{Ott}}, \bibnamefont{and}
  \bibinfo{author}{\bibfnamefont{B.~R.} \bibnamefont{Hunt}},
  \bibinfo{journal}{Phys. Rev. E} \textbf{\bibinfo{volume}{69}},
  \bibinfo{pages}{066215} (\bibinfo{year}{2004}).

\bibitem[{\citenamefont{Sun et~al.}(2009)\citenamefont{Sun, Bollt, and
  Nishikawa}}]{Su:Bo:Ni}
\bibinfo{author}{\bibfnamefont{J.}~\bibnamefont{Sun}},
  \bibinfo{author}{\bibfnamefont{E.~M.} \bibnamefont{Bollt}}, \bibnamefont{and}
  \bibinfo{author}{\bibfnamefont{T.}~\bibnamefont{Nishikawa}},
  \bibinfo{journal}{Europhys. Lett.} \textbf{\bibinfo{volume}{85}},
  \bibinfo{pages}{60011} (\bibinfo{year}{2009}).

\bibitem[{\citenamefont{Sorrentino and Porfiri}(2011)}]{SOPO}
\bibinfo{author}{\bibfnamefont{F.}~\bibnamefont{Sorrentino}} \bibnamefont{and}
  \bibinfo{author}{\bibfnamefont{M.}~\bibnamefont{Porfiri}},
  \bibinfo{journal}{Europhys. Lett.} \textbf{\bibinfo{volume}{93}},
  \bibinfo{pages}{50002} (\bibinfo{year}{2011}).

\bibitem[{\citenamefont{Sorrentino and
  Pecora}(2016)}]{sorrentino2016approximate}
\bibinfo{author}{\bibfnamefont{F.}~\bibnamefont{Sorrentino}} \bibnamefont{and}
  \bibinfo{author}{\bibfnamefont{L.}~\bibnamefont{Pecora}},
  \bibinfo{journal}{Chaos: An Interdisciplinary Journal of Nonlinear Science}
  \textbf{\bibinfo{volume}{26}}, \bibinfo{pages}{094823}
  (\bibinfo{year}{2016}).

\bibitem[{\citenamefont{Cho}(2019)}]{cho2019concurrent}
\bibinfo{author}{\bibfnamefont{Y.~S.} \bibnamefont{Cho}},
  \bibinfo{journal}{Physical Review E} \textbf{\bibinfo{volume}{99}},
  \bibinfo{pages}{052215} (\bibinfo{year}{2019}).

\bibitem[{\citenamefont{Acharyya and
  Amritkar}(2012)}]{acharyya2012synchronization}
\bibinfo{author}{\bibfnamefont{S.}~\bibnamefont{Acharyya}} \bibnamefont{and}
  \bibinfo{author}{\bibfnamefont{R.}~\bibnamefont{Amritkar}},
  \bibinfo{journal}{EPL (Europhysics Letters)} \textbf{\bibinfo{volume}{99}},
  \bibinfo{pages}{40005} (\bibinfo{year}{2012}).

\bibitem[{\citenamefont{Acharyya and
  Amritkar}(2015)}]{acharyya2015synchronization}
\bibinfo{author}{\bibfnamefont{S.}~\bibnamefont{Acharyya}} \bibnamefont{and}
  \bibinfo{author}{\bibfnamefont{R.}~\bibnamefont{Amritkar}},
  \bibinfo{journal}{Physical Review E} \textbf{\bibinfo{volume}{92}},
  \bibinfo{pages}{052902} (\bibinfo{year}{2015}).

\bibitem[{\citenamefont{Shapiro}(1979)}]{SBD1}
\bibinfo{author}{\bibfnamefont{H.}~\bibnamefont{Shapiro}},
  \bibinfo{journal}{Linear Algebra and its applications}
  \textbf{\bibinfo{volume}{25}}, \bibinfo{pages}{129} (\bibinfo{year}{1979}).

\bibitem[{\citenamefont{Maehara and Murota}(2011)}]{SBD2}
\bibinfo{author}{\bibfnamefont{T.}~\bibnamefont{Maehara}} \bibnamefont{and}
  \bibinfo{author}{\bibfnamefont{K.}~\bibnamefont{Murota}},
  \bibinfo{journal}{SIAM J. Matrix Anal. Appl.} \textbf{\bibinfo{volume}{33}},
  \bibinfo{pages}{605} (\bibinfo{year}{2011}).

\bibitem[{\citenamefont{Murota et~al.}(2010)\citenamefont{Murota, Kanno,
  Kojima, and Kojima}}]{SBD3}
\bibinfo{author}{\bibfnamefont{K.}~\bibnamefont{Murota}},
  \bibinfo{author}{\bibfnamefont{Y.}~\bibnamefont{Kanno}},
  \bibinfo{author}{\bibfnamefont{M.}~\bibnamefont{Kojima}}, \bibnamefont{and}
  \bibinfo{author}{\bibfnamefont{S.}~\bibnamefont{Kojima}},
  \bibinfo{journal}{Jpn. J. Ind. Appl. Math.} \textbf{\bibinfo{volume}{27}},
  \bibinfo{pages}{125} (\bibinfo{year}{2010}).

\bibitem[{\citenamefont{Irving and Sorrentino}(2012)}]{Ir:SO}
\bibinfo{author}{\bibfnamefont{D.}~\bibnamefont{Irving}} \bibnamefont{and}
  \bibinfo{author}{\bibfnamefont{F.}~\bibnamefont{Sorrentino}},
  \bibinfo{journal}{Physical Review E} \textbf{\bibinfo{volume}{86}},
  \bibinfo{pages}{056102} (\bibinfo{year}{2012}).

\bibitem[{\citenamefont{Zhang and Motter}(2020)}]{zhang2020symmetry}
\bibinfo{author}{\bibfnamefont{Y.}~\bibnamefont{Zhang}} \bibnamefont{and}
  \bibinfo{author}{\bibfnamefont{A.~E.} \bibnamefont{Motter}},
  \bibinfo{journal}{SIAM Review} \textbf{\bibinfo{volume}{62}},
  \bibinfo{pages}{817} (\bibinfo{year}{2020}).

\bibitem[{\citenamefont{Golub and Kahan}(1965)}]{golub1965calculating}
\bibinfo{author}{\bibfnamefont{G.}~\bibnamefont{Golub}} \bibnamefont{and}
  \bibinfo{author}{\bibfnamefont{W.}~\bibnamefont{Kahan}},
  \bibinfo{journal}{Journal of the Society for Industrial and Applied
  Mathematics, Series B: Numerical Analysis} \textbf{\bibinfo{volume}{2}},
  \bibinfo{pages}{205} (\bibinfo{year}{1965}).

\bibitem[{\citenamefont{Liao et~al.}(2007)\citenamefont{Liao, Wang, and
  Yu}}]{liao2007stability}
\bibinfo{author}{\bibfnamefont{X.}~\bibnamefont{Liao}},
  \bibinfo{author}{\bibfnamefont{L.}~\bibnamefont{Wang}}, \bibnamefont{and}
  \bibinfo{author}{\bibfnamefont{P.}~\bibnamefont{Yu}},
  \emph{\bibinfo{title}{Stability of dynamical systems}}
  (\bibinfo{publisher}{Elsevier}, \bibinfo{year}{2007}).

\bibitem[{\citenamefont{Lorenz}(1963)}]{lorenz1963deterministic}
\bibinfo{author}{\bibfnamefont{E.~N.} \bibnamefont{Lorenz}},
  \bibinfo{journal}{Journal of the atmospheric sciences}
  \textbf{\bibinfo{volume}{20}}, \bibinfo{pages}{130} (\bibinfo{year}{1963}).

\end{thebibliography}
\end{document}